%% file: main.tex
\pdfoutput=1
\documentclass[11pt]{article}

\usepackage{aimpaper}
\usepackage{mathtools,bm}
\usepackage{booktabs,tabularx,array,multirow}
\usepackage{placeins}
\usepackage{enumitem}
\usepackage{tikz}
\usepackage{fontawesome5}
\usetikzlibrary{arrows.meta,positioning,calc,fit,backgrounds,decorations.pathreplacing}
\usepackage[most]{tcolorbox}
\usepackage{algorithm}
\usepackage{algpseudocode}
\usepackage[title,titletoc]{appendix}
\usepackage[nameinlink,capitalise]{cleveref}

\setlist{itemsep=2pt,topsep=4pt}
\renewcommand{\arraystretch}{1.3}

\theoremstyle{plain}
\newtheorem{theorem}{Theorem}
\newtheorem{proposition}{Proposition}
\newtheorem{lemma}{Lemma}

\theoremstyle{definition}

\theoremstyle{remark}
\newtheorem{remark}{Remark}
\theoremstyle{plain}
\newtheorem{applemma}{Lemma}[section]
\crefname{assumption}{Assumption}{Assumptions}
\crefname{algorithm}{Algorithm}{Algorithms}
\crefname{applemma}{Lemma}{Lemmas}

\newcommand{\proofref}[1]{\par\noindent{\small\textit{Proof.} See Appendix~\ref{#1}.}\par\smallskip}

\newcommand{\E}{\mathbb{E}}
\newcommand{\Prob}{\mathbb{P}}
\newcommand{\R}{\mathbb{R}}
\newcommand{\Mset}{\mathcal{M}}
\newcommand{\thI}{\theta^{\mathrm{I}}}
\newcommand{\thS}{\theta^{\mathrm{S}}}
\newcommand{\bI}{b^{\mathrm{I}}}
\newcommand{\bS}{b^{\mathrm{S}}}
\newcommand{\sI}{s^{\mathrm{I}}}
\newcommand{\sS}{s^{\mathrm{S}}}
\newcommand{\MC}{\mathit{MC}}
\newcommand{\wfull}{\hat w}

\definecolor{bidblue}{HTML}{2A78D6}
\definecolor{askorange}{HTML}{EB6834}
\definecolor{aqua}{HTML}{1BAF7A}
\definecolor{ink2}{HTML}{52514E}
\definecolor{figyellow}{HTML}{EDA100}

\newtcolorbox{plainbox}[1][In plain terms]{enhanced, breakable, colback=bidblue!4, colframe=bidblue!50,
  boxrule=0.5pt, arc=2pt, left=6pt, right=6pt, top=3pt, bottom=3pt, before skip=8pt, after skip=8pt,
  fonttitle=\bfseries\small, coltitle=bidblue!65!black, colbacktitle=bidblue!10, toptitle=1pt, bottomtitle=1pt,
  title={#1}, fontupper=\small}
\newtcolorbox{problembox}[1][The problem]{enhanced, breakable, colback=askorange!4, colframe=askorange!55,
  boxrule=0.5pt, arc=2pt, left=6pt, right=6pt, top=3pt, bottom=3pt, before skip=8pt, after skip=8pt,
  fonttitle=\bfseries\small, coltitle=askorange!60!black, colbacktitle=askorange!10, toptitle=1pt, bottomtitle=1pt,
  title={#1}, fontupper=\small}



\input{figures/results.tex}
\input{figures/concept.tex}

\title{Spot Markets for AI Inference}
\subtitle{Dynamic Capacity Allocation and Pricing on a Shared Accelerator Fleet}
\shorttitle{Spot Markets for AI Inference}
\hypersetup{pdftitle={Spot Markets for AI Inference: Dynamic Capacity Allocation and Pricing on a Shared Accelerator Fleet}, pdfauthor={Quanyan Zhu}}
\author{Quanyan Zhu}
\affiliation{Tandon School of Engineering, New York University, Brooklyn, NY, USA}
\email{qz494@nyu.edu}
\paperstatus{Preprint}
\date{8 November 2026}

\begin{document}
\maketitle

\input{sections/00_abstract}

\input{sections/01_intro}
\input{sections/02_why}
\input{sections/03_related}
\input{sections/04_model}
\input{sections/05_structure}
\input{sections/06_mpc}
\input{sections/07_numerics}
\input{sections/08_discussion}
\input{sections/09_conclusion}

\begin{appendices}
\input{sections/appendix}
\end{appendices}

\bibliographystyle{plainnat}
\bibliography{refs}

\end{document}

%% file: figures/results.tex
\newcommand{\rDGainMax}{6.1}
\newcommand{\rDGainOne}{6.6}
\newcommand{\rDMaxHours}{8}
\newcommand{\rDMaxMPC}{371.7}
\newcommand{\rDMaxPosted}{350.2}
\newcommand{\rDOneMPC}{363.6}
\newcommand{\rDOnePosted}{341.0}

\newcommand{\rDayEventMPC}{7.4}

\newcommand{\rDayFlashMPC}{38.4}
\newcommand{\rDayNormalDynamicprice}{2.4}
\newcommand{\rDayNormalMPC}{2.5}
\newcommand{\rDecOther}{0.9}
\newcommand{\rDecPenI}{13.3}
\newcommand{\rDecRevI}{2.7}
\newcommand{\rDecRevS}{12.2}
\newcommand{\rDecSwap}{1.6}
\newcommand{\rDecTotal}{25.6}

\newcommand{\rDelayMPC}{15}

\newcommand{\rDelayPosted}{11}
\newcommand{\rEvAccMPC}{322}
\newcommand{\rEvAccPre}{500}

\newcommand{\rEvLostPosted}{1.2}
\newcommand{\rEvPmaxLarge}{0.78}

\newcommand{\rEvShedKW}{111}
\newcommand{\rEvShedPostedMean}{0.53}
\newcommand{\rEvSpotMPC}{1.2}
\newcommand{\rEvSpotPreMPC}{6.9}

\newcommand{\rExpiredMPC}{0.07}

\newcommand{\rFlashAccBefore}{216}
\newcommand{\rFlashAccPeak}{360}
\newcommand{\rFlashLostDynPrice}{0.6}
\newcommand{\rFlashLostMPC}{0.01}
\newcommand{\rFlashLostPosted}{0.6}
\newcommand{\rFlashPmaxLarge}{0.78}
\newcommand{\rFlashPmaxMedium}{0.73}
\newcommand{\rFlashPmaxSmall}{0.73}
\newcommand{\rFlashStatic}{224}
\newcommand{\rGainDynamicplacement}{5.2}
\newcommand{\rGainDynamicprice}{2.7}
\newcommand{\rGainMPC}{7.4}
\newcommand{\rGainMyopic}{6.9}

\newcommand{\rGainSeMPC}{0.32}

\newcommand{\rHFour}{6.8}
\newcommand{\rHOne}{5.5}
\newcommand{\rHSixteen}{7.1}

\newcommand{\rLtOne}{7.1}
\newcommand{\rLtThree}{7.0}
\newcommand{\rLtTwo}{7.1}
\newcommand{\rLtZero}{7.2}

\newcommand{\rMissMPC}{0.01}

\newcommand{\rMissPosted}{0.25}
\newcommand{\rNoiseHighMPC}{8.5}
\newcommand{\rNoiseHighMyopic}{7.2}
\newcommand{\rNoiseHighPct}{30}
\newcommand{\rNoiseLowMPC}{7.3}
\newcommand{\rNoiseLowMyopic}{6.8}
\newcommand{\rNoiseLowPct}{5}

\newcommand{\rPninefiveMPC}{51}

\newcommand{\rPninefivePosted}{33}

\newcommand{\rProfitMPC}{369.8}

\newcommand{\rProfitPosted}{344.1}

\newcommand{\rSeeds}{20}

\newcommand{\rShareMPC}{93}

\newcommand{\rSpotTokMPC}{526}
\newcommand{\rSpotTokPosted}{495}

\newcommand{\rSurplusGain}{6.5}

\newcommand{\rSwLoadsOne}{1,098}

\newcommand{\rSwLoadsSixtyFour}{6}
\newcommand{\rSwLoadsZero}{7,414}
\newcommand{\rSwOne}{7.1}

\newcommand{\rSwSixtyFour}{2.7}
\newcommand{\rSwZero}{7.7}

\newcommand{\rWelfareGain}{7.3}

%% file: figures/concept.tex
\newcommand{\Hbase}{16}
\newcommand{\Hhours}{4}
\newcommand{\Kseg}{40}
\newcommand{\Wnom}{380}
\newcommand{\accRent}{2.50}
\newcommand{\arPhi}{0.90}
\newcommand{\arSigma}{10}
\newcommand{\betaMult}{5}
\newcommand{\bfHalfLarge}{4.0}
\newcommand{\bfHalfMedium}{3.6}
\newcommand{\bfHalfSmall}{3.0}

\newcommand{\bfZeroLarge}{7.7}
\newcommand{\bfZeroMedium}{6.9}
\newcommand{\bfZeroSmall}{5.8}
\newcommand{\cbGapPct}{34}
\newcommand{\cbN}{20}

\newcommand{\cutErr}{0.0031}
\newcommand{\cutN}{23}
\newcommand{\deadlineD}{8}
\newcommand{\deadlineMin}{120}
\newcommand{\dfMaxDevTicks}{0.5}
\newcommand{\dfModel}{medium}
\newcommand{\dfPmaxFour}{0.68}
\newcommand{\dfPmaxOne}{0.78}
\newcommand{\dfPmaxSixteen}{0.53}

\newcommand{\eAmp}{0.025}
\newcommand{\eMean}{0.065}
\newcommand{\ellLarge}{0.33}
\newcommand{\ellMedium}{0.10}
\newcommand{\ellSmall}{0.03}

\newcommand{\evCapKW}{236}
\newcommand{\evCapPct}{62}
\newcommand{\evHours}{2}
\newcommand{\evNotice}{2}
\newcommand{\evPrice}{0.80}
\newcommand{\evStart}{17:00}
\newcommand{\flashDecay}{3}
\newcommand{\flashHold}{5}

\newcommand{\flashRamp}{1}
\newcommand{\flashStart}{13:00}
\newcommand{\fleetG}{512}

\newcommand{\gammaMult}{0.5}

\newcommand{\hodPeakTrough}{2.6}

\newcommand{\hyFirstForty}{6}
\newcommand{\hyFirstOne}{1.5}
\newcommand{\hyFirstTen}{3}
\newcommand{\hyPmaxForty}{0.78}
\newcommand{\hyPmaxOne}{0.62}
\newcommand{\hyPmaxTen}{0.68}
\newcommand{\hyRTForty}{480}
\newcommand{\hyRTOne}{12}
\newcommand{\hyRTTen}{120}
\newcommand{\hyStart}{13:00}
\newcommand{\hySurge}{2.5}
\newcommand{\kappaAcc}{1.00}
\newcommand{\kappaUnAcc}{0.50}
\newcommand{\loadLarge}{300}
\newcommand{\loadMedium}{90}
\newcommand{\loadSmall}{30}
\newcommand{\lpMs}{7}
\newcommand{\lpVars}{2,320}
\newcommand{\occMax}{68}
\newcommand{\occMean}{49}
\newcommand{\occMin}{33}
\newcommand{\peakUtil}{69}
\newcommand{\peakUtilSafe}{63}
\newcommand{\priceLarge}{3.00}
\newcommand{\priceMedium}{0.80}
\newcommand{\priceSmall}{0.15}
\newcommand{\pwlErrPct}{0.062}
\newcommand{\rhoLarge}{1.40}
\newcommand{\rhoMedium}{1.51}
\newcommand{\rhoSmall}{1.24}
\newcommand{\sSpot}{10}

\newcommand{\spotHalfShare}{35}
\newcommand{\spotSigma}{10}
\newcommand{\stLossPct}{1.8}
\newcommand{\stLossTightPct}{100}
\newcommand{\stNuMax}{1.51}
\newcommand{\stPeakHour}{22:00}
\newcommand{\stPmaxLarge}{0.60}

\newcommand{\stSpillMedium}{31}
\newcommand{\stSpillMult}{1.5}
\newcommand{\stSpillSmall}{31}
\newcommand{\stTightHours}{4}

\newcommand{\weekend}{0.91}
\newcommand{\wone}{0.45}
\newcommand{\wpark}{0.06}
\newcommand{\wzero}{0.25}

%% file: sections/00_abstract.tex
\begin{abstract}
\noindent Inference providers size their accelerator fleets for the interactive peak and sell spare capacity through fixed discounts, such as batch tiers at half price. However, inference differs from conventional cloud capacity in four respects: capacity depends on the latency tier through batching, moving weights between models is slow and costly, spot demand can wait until its deadline, and power can bind. In this paper, we develop a framework for dynamic capacity allocation and pricing of AI inference on a shared accelerator fleet. We consider a provider that serves several models under a power cap and sells each model in an interactive tier with a speed promise and in a spot tier for preemptible work with a deadline. In each period, the provider chooses how many replicas of each model to keep loaded and what spot price to quote. We show that one interactive token displaces a fixed number $\rho_m>1$ of spot tokens, and that the optimal spot price lies midway between the interactive price and the effective cost of capacity in the deadline window, which is close to the prevailing half price only when the fleet is slack. We further show that a spot tier raises the optimal reliability of the interactive tier, and that placement should follow persistent demand shifts while prices absorb transients. Building on these results, we design a model predictive controller based on a linear program with explicit approximation bounds, which solves each period in milliseconds. Numerical experiments on a simulated week with a measured daily demand profile, a flash crowd, and a grid curtailment show that the controller raises profit by $\rGainMPC\%$ over fixed half-price practice, captures $\rShareMPC\%$ of the gain available with perfect forecasts, keeps the interactive promise through both events, and leaves spot buyers better off.
\end{abstract}

%% file: sections/01_intro.tex
\section{Introduction}\label{sec:intro}

Generative AI is increasingly delivered as a metered service. When a person, an application, or an autonomous agent submits a request to a foundation model, the model produces its response one token at a time on a fleet of specialized accelerators, and the provider charges for the tokens consumed \citep{Zhu2026Tokenomics,Demirer2025Intelligence}. As foundation models become embedded in search, software development, customer service, and multi-step agentic workflows, the production of tokens, known as inference, has emerged as a central economic activity of the AI industry. Inference is also intensive in both capital and energy. The accelerators that perform inference are among the most expensive assets in computing, and the data centers that house them have become a rapidly growing source of electricity demand \citep{Shehabi2024,IEA2025}. Therefore, the allocation and pricing of inference capacity is not only an engineering question of throughput and latency but also an economic question about the use of a scarce, costly, and perishable resource.

Inference capacity differs from a conventional commodity in three respects. First, it is perishable. An accelerator-hour that is not used to generate tokens cannot be stored and sold later, and hence idle capacity is a cost that is never recovered. Second, it is slow to reconfigure. To serve a model, a provider must load the model's weights, often tens to hundreds of gigabytes, onto one or more accelerators, which takes from seconds to minutes and interrupts production on those accelerators \citep{Fu2024ServerlessLLM}. Third, the quantity of capacity depends on the quality of service that it delivers. Serving systems process many requests together in a batch \citep{Yu2022Orca,Kwon2023vLLM}, and a larger batch produces more tokens per second in total but fewer for each request \citep{Pope2023}. As a result, a provider that promises interactive users a fast stream of tokens must keep its batches small and accept a lower throughput than the same hardware could deliver for work that can tolerate delay.

These properties shape how inference is provisioned and sold today. Interactive demand follows the working day of its users. Since weights cannot be loaded quickly enough to meet a sudden surge, providers size their fleets for the busiest hour with a margin for forecast error. Consequently, a substantial part of the fleet idles for much of the day. The major providers have responded by selling deferrable work at a discount. Batch interfaces complete requests within a day at half the standard price, flex tiers offer lower prices in exchange for slower or less reliable service, and some providers have charged less in off-peak hours \citep{OpenAI2026BatchFAQ,Anthropic2026Batch,Google2026GeminiBatch,OpenAI2026Pricing,DeepSeek2026Pricing}. One important feature shared by these offers is that their prices are fixed. In other words, the prices do not respond to whether the fleet is busy or idle, and access is rationed through rate limits and queues rather than through prices.

At the same time, the conditions under which inference capacity is supplied are becoming more variable. The release of a new model can raise demand for it sharply within hours. Agents, evaluation suites, and data pipelines generate large volumes of work that can wait minutes or hours for an answer. Moreover, data centers are increasingly constrained by their connections to the power grid, and grid operators are asking large loads to reduce consumption at short notice \citep{IEA2025,Norris2025}. A fixed discount cannot respond to any of these conditions. When the fleet is slack, a fixed discount may be close to the right price. When the fleet is tight, however, it leaves the scarcity value of capacity unpriced, and when the grid is stressed, it provides no signal that would move flexible work to other hours. Nevertheless, a dynamic alternative is not straightforward to design. The provider must decide which models to keep loaded and at what price to sell spare capacity before it observes the next period's interactive load, and it must honor the deadlines of the work that it has already accepted.

Infrastructure clouds faced a related problem when they began to sell idle virtual machines. Spot instances, introduced in 2009, offered spare capacity at a price that varied with its availability \citep{BenYehuda2013Spot,AWS2017Spot}, and a body of work has studied how a data-center operator can adapt its capacity, and in spot markets also its prices, to changing demand \citep{Kusic2009,Zhang2011Spot,Lin2013RightSizing}. However, inference differs from virtual machines in ways that change the structure of the control problem. Two of these differences have been noted above. The capacity of a replica depends on the latency tier that it serves, and moving accelerators between models is slow and costly, even though dividing a model's replicas between tiers is a free setting of the batch scheduler. Two further differences arise from the nature of demand and supply. Spot work has a deadline rather than a start time, so demand can wait while capacity cannot. In addition, power, not only the number of accelerators, can limit production. \Cref{sec:why} explains each of these features in more detail.

In this paper, we develop a framework for dynamic capacity allocation and pricing of AI inference on a shared accelerator fleet. We consider a provider that serves several models under a power cap and sells each model in two tiers: an interactive tier, at a posted price with a guaranteed decode speed, and a spot tier, for preemptible work that must be completed within a deadline, at a price that the provider may revise every period. In each period, the provider decides how many replicas of each model to keep loaded and what spot price to quote before the period's interactive load is realized. The resulting problem is a constrained stochastic control problem with two time scales, a queue with deadlines, and coupling across models through the shared fleet and the power cap. We first study simplified versions of the problem that admit closed-form characterizations, and we then solve the full problem by model predictive control. \Cref{fig:problem} illustrates the setting and its translation into control-theoretic terms.

The contributions of this paper are fourfold. First, we show that continuous batching induces a \emph{latency exchange rate} between the two tiers. At the margin, one interactive token per second displaces $\rho_m>1$ spot tokens per second, where $\rho_m$ is the ratio of a replica's full-load throughputs at the two tiers' batch caps (\cref{lem:cobatch}). This result converts the latency premium of interactive service into a linear capacity constraint that can be priced.

Second, we characterize optimal spot prices. On a shared fleet, the optimal spot price lies midway between the interactive price and the full marginal cost of a spot token, which includes the shadow rents of the fleet and of power. These rents are determined by water-filling across models, and hence a demand surge on one model raises the spot prices of all models (\cref{prop:static}). When the fleet is slack, the rule reproduces the half-price discount observed in practice, and a fixed discount loses little. When the fleet is tight, however, a fixed discount forgoes the scarcity rent (\cref{prop:value}). When work can be deferred, the relevant cost is that of the cheapest capacity within the deadline window (\cref{thm:defer}).

Third, we establish two structural results on reliability and placement. A spot tier acts as a salvage market for the safety capacity of the interactive tier and therefore raises the optimal reliability of interactive service (\cref{thm:backfill}). Moreover, placement exhibits an inaction band determined by the round-trip cost of loading and unloading a model. In other words, transient shocks are absorbed by prices, and only persistent shifts justify moving weights (\cref{thm:hyst}).

Fourth, we design a model predictive controller that embeds these results. A piecewise-linear approximation of revenue, tangent cuts on the expected interactive overflow, and a cumulative representation of deadlines reduce each period's planning problem to a linear program with explicit approximation bounds (\cref{lem:fifo,lem:pwl}). The program solves in about $\lpMs$ milliseconds (\cref{sec:mpc}).

We evaluate the controller on a simulated week of a $\fleetG$-accelerator fleet serving three models, with a daily demand profile measured on a public inference marketplace, a flash crowd triggered by a model launch, and a curtailment requested by the grid operator (\cref{sec:numerics}). Relative to a strong version of current practice, the controller raises profit by $\rGainMPC\%$ and captures $\rShareMPC\%$ of the gain available to a controller with perfect forecasts. It keeps the interactive speed promise through both events, serves more spot work, and increases the surplus of spot buyers. Note that the sources of the gain differ across regimes. On ordinary days, the gain comes mainly from prices, whereas during the two events it comes from reallocating accelerators and power across models. Collectively, these results suggest that a spot tier with dynamic prices is more than a channel for selling idle capacity. It also serves as an instrument for managing the reliability of interactive service, absorbing demand shocks, and offering flexibility to the power grid.

The rest of the paper is organized as follows. \Cref{sec:why} describes how inference capacity is produced and sold and states the problem. \Cref{sec:related} reviews related work. \Cref{sec:model} presents the model, and \cref{sec:structure} characterizes optimal policies in simplified settings. \Cref{sec:mpc} develops the model predictive controller, and \cref{sec:numerics} reports the numerical study. \Cref{sec:discussion} discusses implications and limitations, and \cref{sec:conclusion} concludes the paper. Proofs are collected in Appendix~\ref{app:proofs}.

%% file: sections/02_why.tex
\section{Background and problem statement}\label{sec:why}

In this section, we describe how inference capacity is produced and sold, explain why a substantial part of it idles, and identify the features that distinguish its allocation from that of conventional cloud capacity. We conclude the section with a statement of the problem studied in the rest of the paper. \Cref{fig:problem} summarizes the setting pictorially and restates the problem in the vocabulary of stochastic control for readers who are familiar with the methods but new to AI inference.

\input{sections/fig_problem}

\paragraph{Production of inference.} A provider serves a model by loading a copy of its weights, called a \emph{replica}, onto one or more accelerators. A small dense model may fit on a single accelerator, whereas a large mixture-of-experts model may span eight or more. A request is processed in two phases. In the \emph{prefill} phase, the replica reads the prompt and builds a key--value cache that summarizes it. In the \emph{decode} phase, it generates the response one token at a time, and each token requires a pass through the whole model. Note that, for the batch sizes used in interactive service, a decode step is limited by memory bandwidth rather than arithmetic, since it must read the weights and the caches of all sequences in progress \citep{Pope2023}. Therefore, serving systems process many sequences together so that one pass over the weights advances every sequence in the batch by one token. With continuous batching, sequences join and leave the batch at every step \citep{Yu2022Orca}, and paged memory management allows the caches of many sequences to share the accelerator's memory \citep{Kwon2023vLLM}. A larger batch raises the total throughput of a replica but slows every sequence in it, since each step takes longer. In other words, the batch size is an operating point on a trade-off between throughput and latency, and it is bounded above by the memory available for caches. \Cref{sec:fleet} formalizes this trade-off.

\paragraph{Fleet sizing for the peak.} Interactive demand follows the working day of its users. On one public inference marketplace, the hourly token volume over 30 days has a peak-to-trough ratio of $\hodPeakTrough$ across the hours of the day \citep{Qubax2026API}. Since weights cannot be loaded quickly enough to meet a surge, a provider must keep enough replicas loaded to serve interactive traffic at its busiest hour with a margin for forecast error. A fleet sized for the peak of this profile runs at $\peakUtil\%$ utilization on average, and at $\peakUtilSafe\%$ with a 10\% safety margin. In the fleet simulated in \cref{sec:numerics}, mean interactive load occupies $\occMean\%$ of the accelerators on average, ranging between $\occMin\%$ and $\occMax\%$ over the day. The remaining accelerators represent capacity that the provider has already paid for. Note that accelerators are expensive to own or lease, and an accelerator-hour that is not used cannot be recovered later.

\paragraph{Fixed discounts for spare capacity.} The major providers sell deferral through batch interfaces that complete work within a day at half the standard price \citep{OpenAI2026BatchFAQ,Anthropic2026Batch,Google2026GeminiBatch}. Some providers also sell flex tiers, which may be slower or unavailable at busy times, and priority tiers, which cost more \citep{OpenAI2026Pricing,Anthropic2026Pricing}. One provider has charged lower prices in off-peak hours \citep{DeepSeek2026Pricing}. Across the tiered endpoints listed on one large model router on 31 October 2026, the median flex price is $0.50$ of the standard price of the same model and provider, the median priority price is $2.00$, and the median batch price is $0.50$ \citep{OpenRouter2026API}. These ratios are round numbers set by policy. They do not change when the fleet is busy or idle, and batch tiers ration access with rate limits and queues rather than with prices. Moreover, secondary venues on which inference capacity is resold quote prices well below list \citep{Qubax2026,TokensToCash2026}. This situation resembles the early market for cloud capacity before spot instances priced spare virtual machines according to their availability \citep{BenYehuda2013Spot,AWS2017Spot}.

\paragraph{Distinctive features of inference capacity.} Dynamic pricing of spare capacity is well established for virtual machines. However, four features of inference change the control problem.
\begin{enumerate}[leftmargin=*,label=\arabic*.]
\item \emph{Capacity depends on the service tier.} A replica's throughput rises with its batch size, and the latency promise of the interactive tier caps the batch. Hence, the same replica produces more spot tokens than interactive tokens, and one interactive token displaces $\rho_m>1$ spot tokens (\cref{lem:cobatch}). In contrast, a virtual machine is a fixed unit of capacity regardless of what it is sold as.
\item \emph{Placement is slow and costly, while splitting between tiers is free.} Moving an accelerator from one model to another requires loading tens to hundreds of gigabytes of weights, which takes seconds for a small model and minutes for a large one \citep{Fu2024ServerlessLLM}. In contrast, dividing a model's replicas between interactive and spot work is a setting of the batch scheduler \citep{Yu2022Orca,Kwon2023vLLM}. As a result, the control problem has two time scales.
\item \emph{Demand can be stored, but capacity cannot.} Spot work has a deadline rather than a start time, and hence admitted work waits in a backlog. The backlog is a state variable, and the right price for spot work in the current period depends on the capacity that will be free before its deadline.
\item \emph{Power, not only accelerators, can bind.} Inference facilities are increasingly limited by their grid connections, and grid operators are asking large loads to curtail at short notice \citep{Shehabi2024,IEA2025,Norris2025}. Spot work is the part of the load that can be curtailed without breaking a promise, as long as its deadlines can still be met.
\end{enumerate}

\begin{problembox}[The problem]
Consider a provider that serves several models on one fleet of accelerators under a power cap. The provider sells each model in an interactive tier, at a posted price with a speed promise, and in a spot tier, for preemptible work with a deadline, at a price that it may change every period. In every period, it chooses how many replicas of each model to keep loaded and what spot price to quote before it knows the period's interactive load. The problem is to determine how the provider should set placement and prices to maximize profit, keep the interactive promise, and meet spot deadlines, when loading a model takes time and money, spot work can wait, and power may be curtailed.
\end{problembox}

\paragraph{Significance of the problem.} The problem is important for three reasons. First, the stakes are large. Accelerator fleets are among the most expensive assets in computing, and hence a few percentage points of their utilization represent substantial value. Second, the problem is becoming more pressing. Much of the work that agents and data pipelines generate, such as evaluations, synthetic data, document processing, and overnight agent runs, does not need an answer within seconds. In addition, data centers are becoming the loads that grid operators ask to flex \citep{IEA2025,Norris2025}. Third, a spot price that reflects the scarcity of capacity carries information beyond a single transaction. It tells buyers when flexible work is cheap to run, informs the facility of how much load it can shed when the grid is stressed, and provides a reference price on which forward contracts, capacity reservations, and other instruments for inference capacity can be written. In contrast, a fixed discount conveys none of this information.

%% file: sections/fig_problem.tex
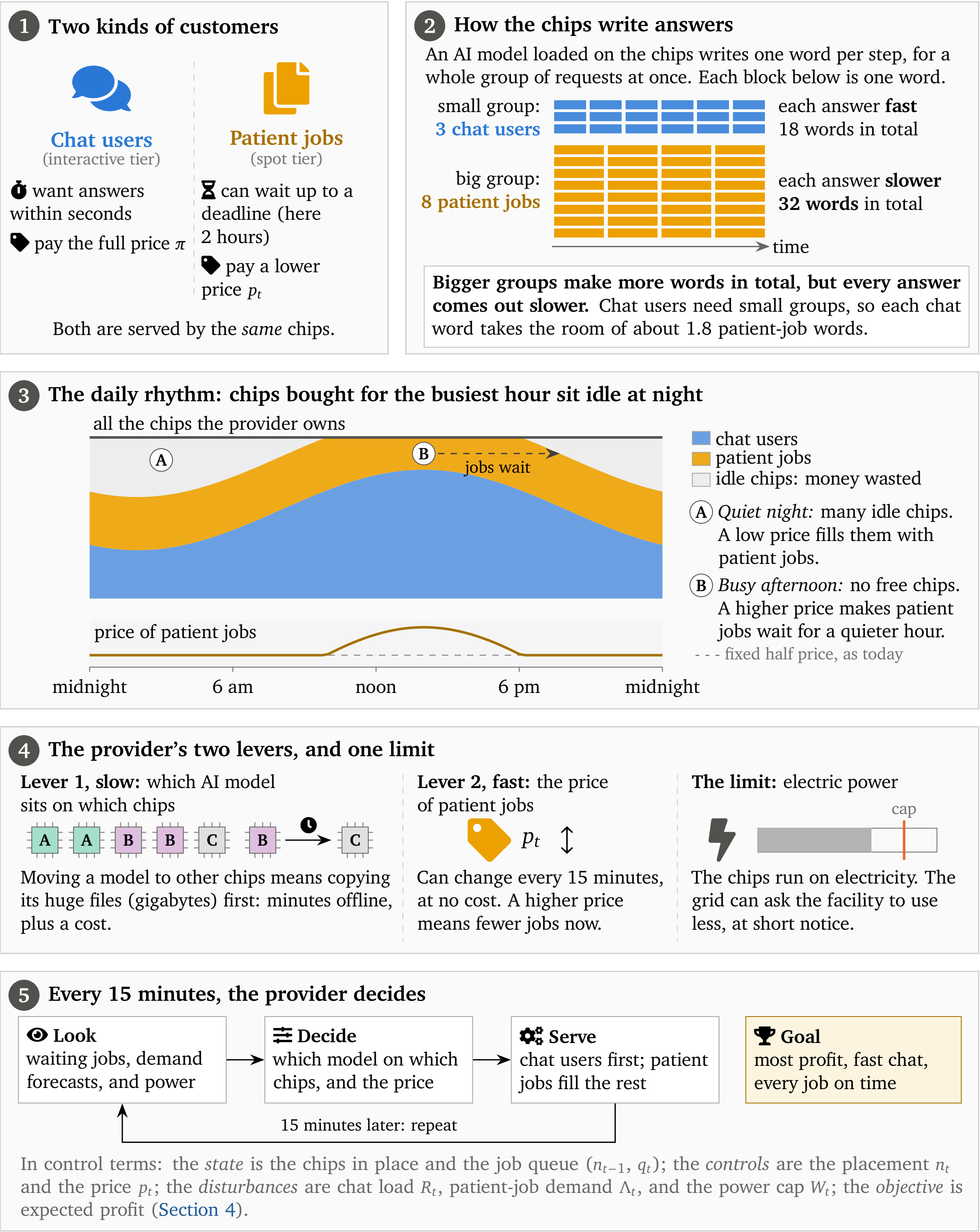
\begin{figure}[p]
\centering
\newcommand{\chip}[4]{
  \foreach \i in {-0.13,0,0.13} {
    \draw[black!45, line width=0.5pt] ({#1+\i},{#2+0.22}) -- ++(0,0.08) ({#1+\i},{#2-0.22}) -- ++(0,-0.08)
      ({#1+0.22},{#2+\i}) -- ++(0.08,0) ({#1-0.22},{#2+\i}) -- ++(-0.08,0);}
  \filldraw[draw=black!60, fill=#3] ({#1-0.22},{#2-0.22}) rectangle ({#1+0.22},{#2+0.22});
  \node[font=\scriptsize\bfseries] at (#1,#2) {#4};}
\colorlet{chatc}{bidblue}
\colorlet{jobc}{figyellow}
\colorlet{jobtxt}{figyellow!70!black}
\colorlet{modA}{aqua!40}
\colorlet{modB}{violet!25}
\colorlet{modC}{black!12}
\begin{tikzpicture}[font=\footnotesize, >={Stealth[length=2.4mm,width=1.8mm]},
  panel/.style={draw=black!18, fill=black!2, line width=0.6pt},
  badge/.style={circle, fill=ink2, text=white, font=\small\bfseries, inner sep=0pt, minimum size=0.52cm},
  head/.style={anchor=west, font=\small\bfseries, inner sep=0pt},
  txt/.style={align=left, inner sep=1pt},
  ctxt/.style={align=center, inner sep=1pt},
  sym/.style={text=black!55, font=\scriptsize, inner sep=1pt},
  callout/.style={draw=black!25, fill=white, align=left, inner sep=4pt},
  marker/.style={circle, draw=ink2, fill=white, font=\scriptsize\bfseries, inner sep=0pt, minimum size=0.38cm},
  loopbox/.style={draw=black!35, fill=white, align=flush left, text width=3.2cm,
                  minimum height=1.45cm, anchor=north west, inner sep=4pt},
  flow/.style={->, semithick},
  declare function={Iload(\h)=0.55+0.25*cos(15*(\h-14)); spotq(\h)=min(0.33,1-Iload(\h));
                    sprice(\h)=0.5+0.9*max(0,Iload(\h)-0.67);}]

\draw[panel] (0,0) rectangle (6.5,-5.9);
\node[badge] at (0.4,-0.4) {1};
\node[head] at (0.8,-0.4) {Two kinds of customers};
\draw[black!20, dashed] (3.25,-1.0) -- (3.25,-5.0);
\node[text=chatc, font=\Huge] at (1.7,-1.45) {\faComments};
\node[ctxt, font=\small\bfseries, text=chatc] at (1.7,-2.3) {Chat users};
\node[sym] at (1.7,-2.65) {(interactive tier)};
\node[txt, align=flush left, text width=2.95cm, anchor=north] at (1.65,-2.95)
  {\faStopwatch\ want answers within seconds\\[3pt] \faTag\ pay the full price~$\pi$};
\node[text=jobc, font=\Huge] at (4.8,-1.45) {\faCopy};
\node[ctxt, font=\small\bfseries, text=jobtxt] at (4.8,-2.3) {Patient jobs};
\node[sym] at (4.8,-2.65) {(spot tier)};
\node[txt, align=flush left, text width=2.95cm, anchor=north] at (4.85,-2.95)
  {\faHourglassHalf\ can wait up to a deadline (here 2~hours)\\[3pt] \faTag\ pay a lower price~$p_t$};
\node[ctxt] at (3.25,-5.5) {Both are served by the \emph{same} chips.};

\draw[panel] (6.8,0) rectangle (16.4,-5.9);
\node[badge] at (7.2,-0.4) {2};
\node[head] at (7.6,-0.4) {How the chips write answers};
\node[txt, align=flush left, anchor=north west, text width=9.2cm] at (7.1,-0.7)
  {An AI model loaded on the chips writes one word per step, for a whole group of requests at once. Each block below is one word.};
\foreach \r in {0,1,2} \foreach \k in {0,...,5}
  \fill[chatc!85] ({9.25+0.6*\k+0.035},{-1.65-0.2*\r}) rectangle ++(0.53,-0.15);
\foreach \r in {0,...,7} \foreach \k in {0,...,3}
  \fill[jobc] ({9.25+0.9*\k+0.035},{-2.4-0.2*\r}) rectangle ++(0.83,-0.15);
\node[txt, anchor=east, align=right] at (9.1,-1.925) {small group:\\\textcolor{chatc}{\textbf{3 chat users}}};
\node[txt, anchor=east, align=right] at (9.1,-3.175) {big group:\\\textcolor{jobtxt}{\textbf{8 patient jobs}}};
\node[txt, anchor=west] at (13.0,-1.925) {each answer \textbf{fast}\\18 words in total};
\node[txt, anchor=west] at (13.0,-3.175) {each answer \textbf{slower}\\\textbf{32 words} in total};
\draw[->, semithick, black!60] (9.25,-4.1) -- (12.9,-4.1) node[right, black, inner sep=1pt] {time};
\node[callout, text width=8.85cm, anchor=north west] at (7.1,-4.42)
  {\textbf{Bigger groups make more words in total, but every answer comes out slower.} Chat users need small groups, so each chat word takes the room of about 1.8 patient-job words.};

\draw[panel] (0,-6.2) rectangle (16.4,-11.85);
\node[badge] at (0.4,-6.6) {3};
\node[head] at (0.8,-6.6) {The daily rhythm: chips bought for the busiest hour sit idle at night};
\begin{scope}[shift={(1.5,-10.0)}, xscale=0.4, yscale=2.7]
  \fill[black!7] (0,0) rectangle (24,1);
  \fill[chatc!70] (0,0) -- plot[domain=0:24, samples=97] (\x,{Iload(\x)}) -- (24,0) -- cycle;
  \fill[jobc!90] plot[domain=0:24, samples=97] (\x,{Iload(\x)})
     -- plot[domain=24:0, samples=97] (\x,{Iload(\x)+spotq(\x)}) -- cycle;
  \draw[very thick, ink2] (0,1) -- (24,1);
  \node[anchor=south west, inner sep=1.5pt] at (0,1) {all the chips the provider owns};
  \node[marker] at (3,0.86) {A};
  \node[marker] at (14,0.9) {B};
  \draw[->, semithick, black!75, dashed] (14.6,0.9) -- (19.7,0.9);
  \node[anchor=north, inner sep=1.5pt, font=\scriptsize] at (17.1,0.88) {jobs wait};
\end{scope}
\begin{scope}[shift={(1.5,-12.95)}, xscale=0.4, yscale=4]
  \fill[black!4] (0,0.45) rectangle (24,0.65);
  \draw[dashed, black!55] (0,0.5) -- (24,0.5);
  \draw[very thick, jobtxt] plot[domain=0:24, samples=97] (\x,{sprice(\x)});
  \node[anchor=north west, inner sep=2pt] at (0,0.65) {price of patient jobs};
  \draw[black!60] (0,0.45) -- (24,0.45);
  \foreach \h/\lab in {0/midnight, 6/6 am, 12/noon, 18/6 pm, 24/midnight} {
    \draw[black!60] (\h,0.45) -- (\h,0.43);
    \node[anchor=north, inner sep=1.5pt, text height=1.6ex, text depth=0.3ex] at (\h,0.43) {\lab};
  }
\end{scope}
\fill[chatc!70] (11.6,-7.2) rectangle ++(0.3,-0.22); \draw[black!35] (11.6,-7.2) rectangle ++(0.3,-0.22);    \node[txt, anchor=west] at (11.95,-7.31) {chat users};
\fill[jobc!90] (11.6,-7.55) rectangle ++(0.3,-0.22); \draw[black!35] (11.6,-7.55) rectangle ++(0.3,-0.22);    \node[txt, anchor=west] at (11.95,-7.66) {patient jobs};
\fill[black!7] (11.6,-7.9) rectangle ++(0.3,-0.22); \draw[black!35] (11.6,-7.9) rectangle ++(0.3,-0.22);     \node[txt, anchor=west] at (11.95,-8.01) {idle chips: money wasted};
\node[marker] at (11.75,-8.55) {A};
\node[txt, align=flush left, anchor=north west, text width=4.1cm] at (12.0,-8.37)
  {\emph{Quiet night:} many idle chips. A low price fills them with patient jobs.};
\node[marker] at (11.75,-9.78) {B};
\node[txt, align=flush left, anchor=north west, text width=4.1cm] at (12.0,-9.6)
  {\emph{Busy afternoon:} no free chips. A higher price makes patient jobs wait for a quieter hour.};
\node[sym, anchor=west] at (11.6,-10.95) {- - - fixed half price, as today};

\draw[panel] (0,-12.15) rectangle (16.4,-15.95);
\node[badge] at (0.4,-12.55) {4};
\node[head] at (0.8,-12.55) {The provider's two levers, and one limit};
\draw[black!20, dashed] (6.75,-12.95) -- (6.75,-15.75);
\draw[black!20, dashed] (11.35,-12.95) -- (11.35,-15.75);
\node[txt, anchor=north west] at (0.3,-12.9) {\textbf{Lever 1, slow:} which AI model\\sits on which chips};
\chip{0.75}{-14.05}{modA}{A} \chip{1.45}{-14.05}{modA}{A} \chip{2.15}{-14.05}{modB}{B}
\chip{2.85}{-14.05}{modB}{B} \chip{3.55}{-14.05}{modC}{C}
\chip{4.4}{-14.05}{modB}{B}
\draw[flow] (4.78,-14.05) -- (5.55,-14.05);
\node[font=\scriptsize] at (5.17,-13.8) {\faClock};
\chip{5.95}{-14.05}{modC}{C}
\node[txt, align=flush left, anchor=north west, text width=6.2cm] at (0.3,-14.5)
  {Moving a model to other chips means copying its huge files (gigabytes) first: minutes offline, plus a cost.};
\node[txt, anchor=north west] at (6.95,-12.9) {\textbf{Lever 2, fast:} the price\\of patient jobs};
\node[text=jobc, font=\huge] at (8.2,-14.05) {\faTag};
\node[font=\large] at (8.9,-14.05) {$p_t$};
\node[font=\Large] at (9.5,-14.05) {$\updownarrow$};
\node[txt, align=flush left, anchor=north west, text width=4.2cm] at (6.95,-14.5)
  {Can change every 15 minutes, at no cost. A higher price means fewer jobs now.};
\node[txt, anchor=north west] at (11.55,-12.9) {\textbf{The limit:} electric power};
\node[font=\huge, text=ink2] at (12.1,-14.05) {\faBolt};
\draw[black!55] (12.7,-13.85) rectangle (15.7,-14.25);
\fill[black!30] (12.7,-13.85) rectangle (14.6,-14.25);
\draw[very thick, askorange] (15.15,-13.72) -- (15.15,-14.38);
\node[sym, anchor=south] at (15.15,-13.72) {cap};
\node[txt, align=flush left, anchor=north west, text width=4.6cm] at (11.55,-14.5)
  {The chips run on electricity. The grid can ask the facility to use less, at short notice.};

\draw[panel] (0,-16.25) rectangle (16.4,-20.6);
\node[badge] at (0.4,-16.65) {5};
\node[head] at (0.8,-16.65) {Every 15 minutes, the provider decides};
\node[loopbox] (look) at (0.3,-17.0) {\faEye\ \textbf{Look}\\ waiting jobs, demand forecasts, and power};
\node[loopbox] (decide) at (4.43,-17.0) {\faSlidersH\ \textbf{Decide}\\ which model on which chips, and the price};
\node[loopbox] (serve) at (8.56,-17.0) {\faCogs\ \textbf{Serve}\\ chat users first; patient jobs fill the rest};
\draw[flow] (look.east) -- (decide.west);
\draw[flow] (decide.east) -- (serve.west);
\draw[flow] (serve.south) -- ++(0,-0.65) -| node[pos=0.25, above=2pt, inner sep=1.5pt, font=\scriptsize] {15 minutes later: repeat} (look.south);
\node[draw=jobtxt, fill=jobc!12, align=flush left, text width=3.33cm, minimum height=1.45cm, anchor=north west, inner sep=4pt]
  at (12.49,-17.0) {\faTrophy\ \textbf{Goal}\\ most profit, fast chat, every job on time};
\node[anchor=north west, text width=15.8cm, align=left, text=black!60, inner sep=1pt] at (0.3,-19.3)
  {In control terms: the \emph{state} is the chips in place and the job queue ($n_{t-1}$, $q_t$); the \emph{controls} are the placement $n_t$ and the price $p_t$; the \emph{disturbances} are chat load $R_t$, patient-job demand $\Lambda_t$, and the power cap $W_t$; the \emph{objective} is expected profit (\cref{sec:model}).};
\end{tikzpicture}
\caption{An overview of the provider's problem in five steps, intended for readers who are new to AI inference. In the figure, ``words'' refer to tokens and ``chips'' refer to accelerators. Panels 2 and 3 use illustrative numbers, and \cref{sec:model} presents the formal model.}
\label{fig:problem}
\end{figure}

%% file: sections/03_related.tex
\section{Related work}\label{sec:related}

In this section, we review related work on cloud spot markets, language-model serving, revenue management, electricity markets, model predictive control, and AI economics.

\paragraph{Spot markets for cloud capacity.} Spot instances sell the idle capacity of an infrastructure cloud at a price that varies with its availability. \Citet{BenYehuda2013Spot} have found that early spot prices were not always set by an open market, and Amazon Web Services later replaced bidding with smoothed prices \citep{AWS2017Spot}. On the provider's side, our previous work \citep{Zhang2011Spot} has formulated capacity allocation and pricing for the spot markets of an infrastructure cloud as a constrained optimal control problem solved by model predictive control, and \citet{Zhang2013Placement} have studied dynamic service placement across data centers with reconfiguration costs. We adopt the receding-horizon architecture of this line of work but change the plant: capacity depends on the service tier through batching, demand can wait in a backlog with deadlines, capacity changes have explicit costs, and power is a constraint. Our provider sets prices openly as a monopolist of its own spare capacity, and \cref{rem:pricetaker} gives its competitive ask. On the buyer's side of a spot market, \citet{Miao2024SpotServe} serve language models on preemptible instances.

\paragraph{Serving language models.} Continuous batching \citep{Yu2022Orca}, paged key--value caches \citep{Kwon2023vLLM}, chunked prefill \citep{Agrawal2024Sarathi}, and disaggregated prefill and decode \citep{Zhong2024DistServe,Patel2024Splitwise} raise the throughput of a replica and make the throughput--latency trade-off explicit, as analyzed by \citet{Pope2023}. Our batching law and the exchange rate $\rho_m$ summarize this trade-off for the controller. AlpaServe \citep{Li2023AlpaServe} places models on a shared cluster to exploit statistical multiplexing, and ServerlessLLM \citep{Fu2024ServerlessLLM} reduces the cost of loading a model. These systems decide placement for a given demand. In contrast, we add prices and a deferrable tier, and we treat loading time and cost as frictions whose economic effect is characterized by \cref{thm:hyst}.

\paragraph{Revenue management and peak-load pricing.} Selling a perishable capacity to price-sensitive demand is the problem of revenue management \citep{Gallego1994,Talluri2004}. One key difference in our setting is that the capacity regenerates every period, its size is a decision, and it is shared across products at a rate that depends on the product. Peak-load pricing \citep{Steiner1957,Crew1995PeakLoad} explains why the price of capacity should rise when it is scarce. \Cref{prop:static} is a monopoly version with several products sharing one plant, and \cref{thm:defer} adds storable demand. The backfill theorem is a newsvendor result \citep{Arrow1951} in which the spot tier serves as the salvage market for safety capacity.

\paragraph{Electricity markets and data-center flexibility.} Spot pricing of electricity over space and time \citep{Bohn1984Spot,Stoft2002,Wilson2002} provides the model for pricing a non-storable good from a constrained network. Data centers can shift workload in time to follow electricity prices or carbon intensity \citep{Qureshi2009Electric,Wiesner2021Wait,Radovanovic2023} and reduce their demand during grid peaks \citep{Liu2013DR}. Dynamic right-sizing turns servers off when load is low and faces the same switching costs as our placement problem \citep{Lin2013RightSizing}. Lookahead control of data centers goes back at least to \citet{Kusic2009}. Our contribution to this line of work is to price the flexibility. The spot tier is the curtailable part of the load, and its price tells buyers when curtailment is likely.

\paragraph{Model predictive control.} Receding-horizon control of constrained systems has a mature theory \citep{Mayne2000,Rawlings2017}. We use it as a practical method and do not claim stability or optimality guarantees for the closed loop. Instead, the numerical study measures how close it comes to a controller with perfect forecasts. Note that the inaction bands of \cref{thm:hyst} are the discrete-time analog of those in investment under adjustment costs \citep{Dixit1989,Eberly1997}.

\paragraph{The economics of AI inference.} \Citet{Demirer2025Intelligence} document the pricing, supply, and demand of language-model inference, and in \citet{Zhu2026Tokenomics}, we have studied the economics of tokens, computation, and pricing in foundation models. These studies take the provider's capacity and costs as given or focus on posted prices. In contrast, this paper takes the provider's side. We derive the cost of capacity from the fleet, its batching, and its power supply, and we show what spot price a provider should quote as a result.

%% file: sections/04_model.tex
\section{The model}\label{sec:model}

In this section, we present the model of the fleet, the demand, and the dynamics, and we formulate the provider's control problem. We divide time into control periods $t=0,1,\dots$ of length $\Delta$ seconds, and we set $\Delta=15$ minutes in the numerical study. Within a period, the set of loaded replicas is fixed, but the scheduler can change, at every decode step, which sequences share a batch. Therefore, the model has two layers: a slow layer of \emph{placement}, decided at the start of each period, and a fast layer of \emph{batching} inside the period, which we summarize by its throughput.

Placement and batching correspond to layers 2 and 3 of \cref{fig:layers}, which places them within a broader view of the provider's operations, from the hardware at the bottom to the buyers at the top. Capacity is built up from the accelerators through the deployments, the serving engine, and the queues, and it is sold as products and tiers to buyers whose demand responds to price and latency. For each layer, the figure indicates its time scale and the elements of the model that represent it. Two features of the figure deserve emphasis. First, changing a deployment, which requires loading weights and warming caches, is a transition within layer 2 rather than a layer of its own; the model captures it through the loading time $L_m$ and the costs $\kappa^\pm_m$. Second, the controller acts on only two layers, the placement and the spot price, and it observes the others.

\input{sections/fig_layers}

\subsection{Fleet, models, and batching}\label{sec:fleet}

Consider a provider that operates a fleet of $G$ identical accelerators in one facility with a power cap of $W_t$ kilowatts. The provider serves a set $\Mset=\{1,\dots,M\}$ of models. A \emph{replica} of model $m$ refers to one loaded copy of its weights, and it occupies $g_m$ accelerators (e.g., one accelerator for a small dense model and eight for a large mixture-of-experts model).

\paragraph{Decode.} We adopt a first-order model of memory-bound decode \citep{Pope2023}. In each step, a replica reads its weights once and the key--value cache of every sequence in its batch. Hence, a decode step of a replica that holds a batch of $b$ sequences takes
\begin{equation}\label{eq:tau}
\tau_m(b)=\tau_{0,m}+\tau_{1,m}\,b,\qquad 0\le b\le B_m,
\end{equation}
where $\tau_{0,m}$ denotes the time to stream the weights, $\tau_{1,m}$ the per-sequence cost of attention and cache traffic, and $B_m$ the largest batch whose key--value caches fit in memory. Since every sequence in the batch advances one token per step, each stream runs at $1/\tau_m(b)$ tokens per second, and the replica produces $\theta_m(b)=b/\tau_m(b)$ tokens per second. Note that the throughput $\theta_m$ is increasing and concave in $b$.

\paragraph{Two tiers.} The provider sells two service tiers for each model.
\begin{enumerate}[label=(\arabic*),leftmargin=*]
\item The \emph{interactive} tier promises every stream a decode speed of at least $\sI_m$ tokens per second. It is sold at a posted price $\pi_m$ per token, and its demand is exogenous in the short run.
\item The \emph{spot} tier promises completion within a deadline and a much lower speed floor $\sS<\sI_m$ (or none). Spot work is preemptible, i.e., the provider may pause it at any step. Its price $p_{m,t}$ is set every period.
\end{enumerate}
The speed floors cap the batch. A replica that holds any interactive stream can run batches of at most $\bI_m=\min\{B_m,(1/\sI_m-\tau_{0,m})/\tau_{1,m}\}$ sequences, whereas a replica that holds only spot work can run batches of up to $\bS_m=\min\{B_m,(1/\sS-\tau_{0,m})/\tau_{1,m}\}$. We write
\[
\thI_m=\theta_m(\bI_m),\qquad \thS_m=\theta_m(\bS_m),\qquad \rho_m=\thS_m/\thI_m\ \ge 1
\]
for the full-load throughputs of a replica in the two modes and their ratio. We assume that $\sI_m<1/\tau_{0,m}$ so that interactive service is feasible.

\begin{lemma}[Co-batching and the latency exchange rate]\label{lem:cobatch}
Assume that replicas are divisible. Let $n>0$ replicas of model $m$ carry an interactive load of $r\in[0,\thI_m n]$ tokens per second. Then, the largest spot throughput compatible with the interactive speed floor is
\begin{equation}\label{eq:cap}
\kappa_m(n,r)=\thS_m\,n-\rho_m\,r .
\end{equation}
It is attained by running $r/\thI_m$ replicas at batch $\bI_m$ with interactive streams only and the remaining $n-r/\thI_m$ replicas at batch $\bS_m$ with spot streams only. With whole replicas, running $k=\lceil r/\thI_m\rceil$ replicas at batch $\bI_m$ and filling their spare slots with spot streams yields a spot throughput of at least $\kappa_m(n,r)-(\thS_m-\thI_m)$.
\end{lemma}
\proofref{app:pf-cobatch}

\begin{figure}[t]
\centering
\includegraphics[width=\textwidth]{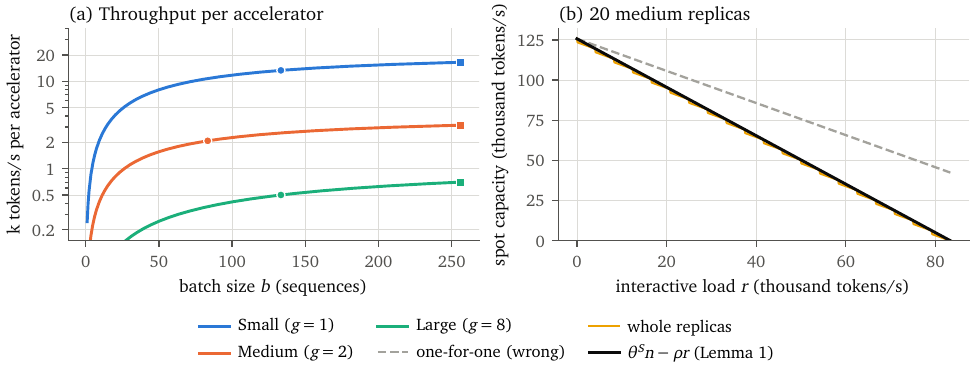}
\caption{Batching and the latency exchange rate. (a) Decode throughput per accelerator, $\theta_m(b)/g_m$, of the three models of \cref{tab:params}. Circles mark the interactive batch cap $\bI_m$ set by the speed floor, and squares mark the spot batch cap $\bS_m$ set by memory. (b) Spot capacity of $\cbN$ medium replicas as their interactive load $r$ grows. The line $\thS n-\rho r$ of \cref{lem:cobatch} is attainable with whole replicas up to the sawtooth shown. A one-for-one rule, which treats an interactive token as displacing one spot token, overstates the largest interactive load that the replicas can carry by $\cbGapPct\%$.}
\label{fig:batch}
\end{figure}

\Cref{lem:cobatch} converts the latency premium of interactive service into an exchange rate between the tiers. In other words, at the margin, one interactive token per second displaces $\rho_m$ spot tokens per second. When neither batch cap is set by memory, $\rho_m=(1-\tau_{0,m}\sS)/(1-\tau_{0,m}\sI_m)$. In the calibration of \cref{sec:numerics}, the spot batch is set by memory, and $\rho_m$ is \rhoSmall, \rhoMedium{}, and \rhoLarge{} for the small, medium, and large models, respectively (\cref{tab:params,fig:batch}). The lemma also shows that the split of a model's replicas between the tiers costs nothing to change. This split is a scheduler setting rather than a placement decision, and hence the controller can leave it to the batch scheduler inside the period.

\subsection{Demand}\label{sec:demand}

\paragraph{Interactive load.} Let $R_{m,t}$ denote the average rate, in tokens per second, at which interactive traffic for model $m$ arrives during period $t$. This rate is random, and the provider forecasts it with mean $\hat r_{m,t}$ and standard deviation $\sigma_{m,t}$ (\cref{sec:forecast}). Each interactive token that is served earns $\pi_m$. A token that cannot be served at the promised speed is lost and incurs a penalty $\beta_m$ in service credits and goodwill.

\paragraph{Spot demand.} In period $t$, a population of spot buyers for model $m$ would buy $\Lambda_{m,t}$ tokens of work at a price of zero. Their valuations per token are uniform on $[0,\pi_m]$, so that no buyer pays more for deferrable service than the interactive price at which it could buy immediate service. At spot price $p$, the work admitted is
\begin{equation}\label{eq:demand}
a_{m,t}=\Lambda_{m,t}\,(1-p/\pi_m)^+,\qquad\text{equivalently}\qquad p=\pi_m\,(1-a_{m,t}/\Lambda_{m,t}).
\end{equation}
Linear demand is a common choice in the pricing of cloud capacity \citep{Zhang2011Spot}. It keeps the pricing rules in closed form, and its slope $\alpha_{m,t}=\Lambda_{m,t}/\pi_m$ scales with the size of the buyer population. Note that spot revenue $a\,\pi_m(1-a/\Lambda_{m,t})$ is a concave quadratic in the admitted volume, while every constraint of the model is linear in it. Therefore, we optimize over admissions and recover the price from the demand curve.

\paragraph{Deadlines.} Work admitted in period $t$ must be completed by the end of period $t+D-1$. Work that misses its deadline expires, its price is refunded, and the provider pays a penalty $\gamma_m$ per token. This convention matches current batch interfaces, which do not charge for requests that expire \citep{OpenAI2026BatchFAQ}. While admitted work waits, the provider incurs a small holding cost $h_m$ per token per period, which captures the buyers' preference for earlier completion and breaks ties in favor of serving early. The deadline $D$ is a contract term. For example, $D=1$ corresponds to a spot tier with no deferral, $D=8$ (two hours) is in the range of current flex tiers, and $D=96$ (a day) corresponds to a batch tier.

\subsection{Dynamics, power, and costs}\label{sec:dynamics}

\paragraph{Placement.} Let $n_{m,t}\ge0$ denote the number of replicas of model $m$ loaded in period $t$. These replicas must fit in the fleet:
\begin{equation}\label{eq:fleet}
\textstyle\sum_{m}g_m\,n_{m,t}\le G .
\end{equation}
Loading a replica takes $L_m$ seconds of weight transfer and warm-up, during which its accelerators produce nothing. Let $u_{m,t}=(n_{m,t}-n_{m,t-1})^+$ denote the loads and $v_{m,t}=(n_{m,t-1}-n_{m,t})^+$ the unloads. Then, the replicas that are \emph{effective} in period $t$ are
\begin{equation}\label{eq:neff}
\tilde n_{m,t}=n_{m,t}-\ell_m\,u_{m,t},\qquad \ell_m=L_m/\Delta .
\end{equation}
Each load also incurs a direct cost $\kappa^+_m$, and each unload incurs a cost $\kappa^-_m$ for the in-flight sequences that it drains and the caches that it discards. Accelerators that hold no model are parked.

\paragraph{Service.} Interactive traffic has priority. The provider serves $Y_{m,t}=\min\{R_{m,t},\thI_m\tilde n_{m,t}\}$ tokens per second, and the overflow $O_{m,t}=R_{m,t}-Y_{m,t}$ is lost. By \cref{lem:cobatch}, the remaining spot capacity of the period is
\begin{equation}\label{eq:spotcap}
K_{m,t}=\Delta\,\bigl(\thS_m\tilde n_{m,t}-\rho_m Y_{m,t}\bigr)\ \ \text{tokens}.
\end{equation}
Spot work is served in first-in-first-out order. Let $q_{m,t}$ denote the backlog at the start of period $t$, $x_{m,t}\le K_{m,t}$ the spot work served, and $z_{m,t}$ the work that expires. Then,
\begin{equation}\label{eq:backlog}
q_{m,t+1}=q_{m,t}+a_{m,t}-x_{m,t}-z_{m,t}.
\end{equation}
Since all spot work of model $m$ has the same relative deadline, first-in-first-out is also earliest-deadline-first. This property allows us to write the deadlines as linear constraints on cumulative service (\cref{lem:fifo}).

\paragraph{Power and energy.} A loaded but idle accelerator draws $w^0$ kW, a parked accelerator draws $w^{\mathrm{park}}<w^0$, and decode at full load adds $w^1$. Power rises in proportion to the share of a replica's capacity in use, and hence the facility draws
\begin{equation}\label{eq:power}
\Omega_t=w^{\mathrm{park}}G+\sum_m g_m\Bigl[(w^0-w^{\mathrm{park}})\,n_{m,t}+w^1\Bigl(\frac{Y_{m,t}}{\thI_m}+\frac{x_{m,t}}{\Delta\thS_m}\Bigr)\Bigr]\ \le W_t .
\end{equation}
Note that a replica at full load draws $g_m(w^0+w^1)$ in either mode. Electricity costs $e_t$ dollars per kWh. Both the price $e_t$ and the cap $W_t$ may change at short notice (e.g., when the grid operator calls a demand-response event).

\paragraph{Profit.} The profit of period $t$ is given by
\begin{equation}\label{eq:profit}
J_t=\sum_m\Bigl[\Delta\pi_m Y_{m,t}+p_{m,t}a_{m,t}-\text{refunds}_{m,t}-\gamma_m z_{m,t}-\Delta\beta_m O_{m,t}-h_m q_{m,t+1}-\kappa^+_m u_{m,t}-\kappa^-_m v_{m,t}\Bigr]-\frac{\Delta}{3600}\,e_t\,\Omega_t ,
\end{equation}
where refunds are the prices paid for the expiring work $z_{m,t}$. Since accelerators are owned or leased for much longer than a period, their cost is sunk and does not appear in \eqref{eq:profit}. Instead, it enters the analysis through the shadow price of the fleet constraint \eqref{eq:fleet}.

\subsection{The control problem}\label{sec:control}

At the start of period $t$, the provider knows the history up to $t-1$, namely, past loads, the backlog with its deadlines, and the replicas in place. It also knows the announced electricity prices and power caps. The provider then chooses the placement $n_t=(n_{m,t})_m$ and the spot prices $p_t=(p_{m,t})_m$. Afterwards, interactive load $R_t$ and the spot intercepts $\Lambda_t$ are realized, interactive traffic is served, and the spot backlog is served within the remaining capacity and power. The provider's problem is formulated as
\begin{equation*}\label{eq:P}\tag{P}
\max_{\text{non-anticipative }(n_t,p_t)}\ \E\Bigl[\sum_{t=0}^{T-1}J_t\Bigr]\quad\text{subject to \eqref{eq:demand}--\eqref{eq:power}}.
\end{equation*}
\Cref{fig:loop} illustrates the feedback loop. Problem \eqref{eq:P} is a constrained stochastic control problem. Its state, which includes the replicas, the backlog by deadline, and the forecast state, is too large for exact dynamic programming. Moreover, its controls interact through the shared fleet and power cap, and its dynamics have two time scales. Therefore, \cref{sec:structure} first studies deterministic and one-period versions of the problem to learn what optimal policies look like, and \cref{sec:mpc} then solves \eqref{eq:P} by model predictive control.

\begin{figure}[t]
\centering
\begin{tikzpicture}[font=\footnotesize, >={Stealth[length=2.4mm,width=1.8mm]},
  box/.style={draw=black!60, fill=black!6, align=center, inner xsep=4pt, inner ysep=4pt},
  ctl/.style={box, draw=askorange!85!black, fill=askorange!10},
  env/.style={box, draw=bidblue!85!black, fill=bidblue!10},
  row1/.style={minimum height=22mm},
  row2/.style={minimum height=18mm},
  flow/.style={->, semithick, black!70}]
\node[ctl, row1, text width=36mm] (mpc) at (0,0) {\textbf{Controller (MPC)}\\ forecasts $\hat r,\sigma,\hat\Lambda$; solves an LP over $H$ periods; applies the first step};
\node[box, row1, text width=40mm] (fleet) at (6.38,0) {\textbf{Fleet}\\ $G$ accelerators, power cap $W_t$\\ replicas $n_{m,t}$, loads cost $\ell_m,\kappa^\pm_m$};
\node[box, row2, text width=40mm] (sched) at (6.38,-2.7) {\textbf{Batch scheduler}\\ interactive first at batch $\le\bI_m$;\\ spot fills the rest (\cref{lem:cobatch})};
\node[env, row1, text width=34mm] (int) at (11.66,0) {\textbf{Interactive buyers}\\ load $R_{m,t}$ at price $\pi_m$\\ speed floor $\sI_m$};
\node[env, row2, text width=34mm] (spot) at (11.66,-2.7) {\textbf{Spot buyers}\\ admit $a_{m,t}$ at price $p_{m,t}$\\ deadline $D$ periods};
\node[box, row2, text width=36mm] (q) at (0,-2.7) {\textbf{Spot backlog}\\ FIFO queue $q_{m,t}$ with deadlines};
\draw[flow] (mpc) -- node[above=2pt] {placement $n_t$} (fleet);
\draw[flow] (fleet) -- (sched);
\draw[flow] (int) -- (fleet);
\draw[flow] (spot) -- (sched);
\draw[flow] (mpc.north) -- (0,1.5) -- node[above=2pt] {spot prices $p_t$} (14.15,1.5) |- (spot.east);
\draw[flow] (sched) -- node[above=2pt] {served $x_{m,t}$} (q);
\draw[flow] (q) -- node[left=2pt] {state} (mpc);
\end{tikzpicture}
\caption{The control loop. Once a period, the controller chooses how many replicas of each model to keep loaded and what spot price to quote. Within the period, the batch scheduler serves interactive traffic first and fills the remaining capacity with spot work.}
\label{fig:loop}
\end{figure}

%% file: sections/fig_layers.tex
\begin{figure}[t]
\centering
\newcommand{\tscale}{\scriptsize\itshape\color{black!55}}
\newcommand{\hdr}{\scriptsize\bfseries\color{black!60}}
\newcolumntype{I}{>{\centering\arraybackslash}m{0.6cm}}
\newcolumntype{T}[1]{>{\raggedright\hyphenpenalty=10000\exhyphenpenalty=10000\arraybackslash}m{#1}}
\newcolumntype{Z}{@{\hspace{2pt}}IT{2.7cm}T{4.0cm}T{3.45cm}@{}}
\newcommand{\layer}[5]{\setlength{\tabcolsep}{3pt}\renewcommand{\arraystretch}{1}%
  \begin{tabular}{Z}{\Large\color{ink2}#1}\newline{\scriptsize\bfseries #2} & #3 & #4 & #5\end{tabular}}
\begin{tikzpicture}[font=\footnotesize, >={Latex[length=1.6mm]},
  band/.style={anchor=north west, inner xsep=0.1cm, inner ysep=3pt, minimum height=1.2cm},
  supply/.style={band, draw=black!30, fill=black!3},
  market/.style={band, draw=askorange!60, fill=askorange!6},
  ctl/.style={anchor=west, align=flush left, text width=3.2cm, inner sep=0pt},
  lever/.style={->, very thick, bidblue!80!black},
  info/.style={->, black!50}]

\node[market] (L6) at (0.85,0) {\layer{\faUsers}{6}
  {\textbf{Demand}\newline{\tscale seconds to days}}
  {Buyers respond to price and latency, and they may switch between models.}
  {interactive load $R_{m,t}$\newline spot: $\Lambda_{m,t}(1-p/\pi_m)^+$\newline\textit{switching: not modeled}}};
\node[market] (L5) at ($(L6.south west)+(0,-0.12)$) {\layer{\faTags}{5}
  {\textbf{Products and tiers}\newline{\tscale posted price: months;\newline spot price: 15 min}}
  {Models of several sizes and grades, each sold in an interactive and a spot tier.}
  {models $m\in\Mset$\newline interactive: $\pi_m$, floor $\sI_m$\newline spot: $p_{m,t}$, deadline $D$}};
\node[supply] (L4) at ($(L5.south west)+(0,-0.12)$) {\layer{\faStream}{4}
  {\textbf{Queues and deferred work}\newline{\tscale minutes to hours}}
  {Interactive requests go first; spot work waits but must finish by its deadline.}
  {served $Y_{m,t}$ and $x_{m,t}$\newline backlog $q_{m,t}$ (FIFO)\newline deadlines: \cref{lem:fifo}}};
\node[supply] (L3) at ($(L4.south west)+(0,-0.12)$) {\layer{\faCogs}{3}
  {\textbf{Serving engine}\newline{\tscale milliseconds}}
  {Continuous batching, with a measured throughput for each model on its hardware.}
  {step time $\tau_m(b)$\newline throughputs $\thI_m$ and $\thS_m$\newline $\rho_m=\thS_m/\thI_m$ (\cref{lem:cobatch})}};
\node[supply] (L2) at ($(L3.south west)+(0,-0.12)$) {\layer{\faServer}{2}
  {\textbf{Deployments}\newline{\tscale minutes}}
  {Replicas of each model on $g_m$ accelerators; a change loads weights and warms caches.}
  {replicas $n_{m,t}$\newline fleet: $\sum_mg_mn_{m,t}\le G$\newline load time $L_m$, costs $\kappa^\pm_m$}};
\node[supply] (L1) at ($(L2.south west)+(0,-0.12)$) {\layer{\faMicrochip}{1}
  {\textbf{Hardware and power}\newline{\tscale fleet: years; cap: hours}}
  {Identical accelerators in one facility, under a power cap that the grid may lower.}
  {$G$ accelerators\newline power cap $W_t$, price $e_t$\newline power draw $w^0$ and $w^1$}};

\coordinate (cx) at ($(L6.east)+(0.4,0)$);
\node[anchor=south west, inner sep=0pt] at ($(L6.north west)+(0.1,0.08)$)
  {\setlength{\tabcolsep}{3pt}\begin{tabular}{Z}& \hdr Layer (time scale) & \hdr What happens there & \hdr In the model\end{tabular}};
\node[anchor=south west, inner sep=0pt] at ($(L6.north -| cx)+(0,0.08)$)
  {\setlength{\tabcolsep}{0pt}\begin{tabular}{l}\hdr Controller, every 15 min\end{tabular}};

\coordinate (ax) at (0.25,0);
\coordinate (dx) at (0.6,0);
\draw[->, thick, ink2] (L1.south -| ax) -- node[pos=0.28, rotate=90, fill=white, inner sep=1.5pt, text=ink2] {capacity is built up} (L6.north -| ax);
\draw[->, thick, askorange!80!black] (L6.north -| dx) -- node[pos=0.28, rotate=90, fill=white, inner sep=1.5pt, text=askorange!80!black] {demand reaches down} (L1.south -| dx);

\draw[draw=bidblue!60!black, fill=bidblue!5] (L6.north -| cx) rectangle ($(L1.south -| cx)+(3.42,0)$);
\draw[info]  (L6.east) -- (L6.east -| cx);
\node[ctl] at ($(L6.east -| cx)+(0.13,0)$) {forecasts the demand: $\hat r_{m,t}$, $\sigma_{m,t}$, and $\hat\Lambda_{m,t}$};
\draw[lever] (L5.east -| cx) -- (L5.east);
\node[ctl] at ($(L5.east -| cx)+(0.13,0)$) {\textbf{fast lever:} posts the spot price $p_{m,t}$};
\draw[info]  (L4.east) -- (L4.east -| cx);
\node[ctl] at ($(L4.east -| cx)+(0.13,0)$) {observes the backlog and its deadlines};
\draw[info]  (L3.east) -- (L3.east -| cx);
\node[ctl] at ($(L3.east -| cx)+(0.13,0)$) {uses the measured throughputs as parameters};
\draw[lever] (L2.east -| cx) -- (L2.east);
\node[ctl] at ($(L2.east -| cx)+(0.13,0)$) {\textbf{slow lever:} sets the placement $n_{m,t}$};
\draw[info]  (L1.east) -- (L1.east -| cx);
\node[ctl] at ($(L1.east -| cx)+(0.13,0)$) {reads the announced cap $W_t$ and price $e_t$};
\end{tikzpicture}
\caption{A layered view of inference supply and demand, from the hardware at the bottom to the buyers at the top. Capacity is built up through the four supply layers (gray), and demand reaches down from the two market layers (orange). Each layer is labeled with its time scale and with the elements of the model that represent it. The controller of \cref{sec:mpc} acts on two layers, through the placement (the slow lever) and the spot price (the fast lever), and it observes the others. Changing a deployment is a transition within layer 2 rather than a layer of its own, and switching between models in layer 6 lies outside the scope of the model.}
\label{fig:layers}
\end{figure}

%% file: sections/05_structure.tex
\section{Structure of optimal policies}\label{sec:structure}

In this section, we study four simplified versions of \eqref{eq:P}. Each version isolates one feature of the inference setting, namely, a shared fleet whose capacity depends on the tier, storable demand, random interactive load, and costly placement. The results explain the behavior of the controller of \cref{sec:mpc}, and they provide the provider with pricing rules that can be published.

Throughout this section, we let $\lambda^e_t=e_t\Delta/3600$ denote the cost of drawing one kW for one period, and we let
\[
\wfull=w^0-w^{\mathrm{park}}+w^1
\]
denote the power that a fully loaded accelerator draws in excess of a parked one.

\subsection{One period: shadow prices of a shared fleet}\label{sec:static}

Consider a single period in which interactive load $R_m$ and the spot intercept $\Lambda_m$ are known, spot work must be served within the period ($D=1$), and the previous placement does not matter (no loading cost). The provider chooses placement and spot admissions by solving
\begin{equation}\label{eq:static}
\begin{aligned}
\max_{n,a\ge0}\ &\sum_m\Bigl[a_m\pi_m\Bigl(1-\frac{a_m}{\Lambda_m}\Bigr)-\lambda^e\Bigl(g_m(w^0-w^{\mathrm{park}})n_m+w^1g_m\Bigl(\frac{R_m}{\thI_m}+\frac{a_m}{\Delta\thS_m}\Bigr)\Bigr)\Bigr]\\
\text{s.t. }\ & a_m\le\Delta(\thS_m n_m-\rho_m R_m)\ \ (m\in\Mset),\qquad \sum_m g_mn_m\le G\ \ [\nu],\qquad \Omega\le W\ \ [\eta].
\end{aligned}
\end{equation}
We assume that interactive load alone fits, i.e., $\sum_m g_mR_m/\thI_m\le G$, and that the power it draws, with every replica it needs loaded, is at most $W$. For each model, we define the \emph{full marginal cost of a spot token} as
\begin{equation}\label{eq:MC}
\MC_m(\nu,\eta)=\frac{g_m\bigl[\nu+(\lambda^e+\eta)\,\wfull\bigr]}{\Delta\,\thS_m}.
\end{equation}
In other words, this quantity is the cost of one replica-period, which consists of the fleet's shadow rent $\nu$ per accelerator plus the energy and the shadow price of power at full load, divided by the spot tokens that the replica-period produces.

\begin{proposition}[Spot prices on a shared fleet]\label{prop:static}
Problem \eqref{eq:static} has an optimal solution with multipliers $\nu\ge0$ for the fleet and $\eta\ge0$ for power such that:
\begin{enumerate}[label=(\roman*)]
\item \emph{Spot prices.} Model $m$ sells spot work if and only if $\MC_m<\pi_m$, and then its spot price and volume are
\begin{equation}\label{eq:pstatic}
p_m=\tfrac12\bigl(\pi_m+\MC_m\bigr),\qquad a_m=\tfrac12\Lambda_m\bigl(1-\MC_m/\pi_m\bigr).
\end{equation}
\item \emph{Interactive opportunity cost.} At the margin, an interactive token costs $\rho_m\MC_m$. For every model with $R_m>0$, $-\Delta\rho_m\MC_m$ is a supergradient of the optimal profit with respect to $R_m$: one more token per second of interactive load reduces the period's profit by at least $\Delta\rho_m\MC_m$, and by exactly that wherever the optimal profit is differentiable in $R_m$.
\item \emph{Water-filling.} Suppose that the power cap is slack at the solution described here, so that $\eta=0$. Let
\[
N(\nu)=\sum_m g_m\Bigl[\frac{R_m}{\thI_m}+\frac{\Lambda_m}{2\Delta\thS_m}\Bigl(1-\frac{\MC_m(\nu,0)}{\pi_m}\Bigr)^+\Bigr]
\]
denote the number of accelerators that the provider would use at shadow rent $\nu$. Then, $\nu=0$ if $N(0)\le G$; otherwise, $\nu$ is the unique solution of $N(\nu)=G$.
\item \emph{Spillovers.} When $\eta=0$, $\nu$ is nondecreasing in every $R_{m'}$ and $\Lambda_{m'}$. Hence, every model's spot price is nondecreasing, and its spot volume nonincreasing, in the interactive load and spot demand of every other model.
\item \emph{Slack fleet.} If neither the fleet nor power binds, $p_m=\frac12\pi_m+\frac12 g_m\lambda^e\wfull/(\Delta\thS_m)$: half the interactive price, plus half the energy cost of a spot token.
\end{enumerate}
\end{proposition}
\proofref{app:pf-static}

Part (v) offers an interpretation of current practice. When the fleet is slack, the monopoly price for spare capacity is half the interactive price, up to a negligible energy term (\cref{tab:params}). This is exactly the 50\% discount that current batch tiers charge \citep{OpenAI2026BatchFAQ,Anthropic2026Batch,Google2026GeminiBatch}. Note that the result depends on the assumption that valuations are uniform up to the interactive price, and hence it should not be interpreted as a structural law. Nevertheless, it does indicate where a fixed half-price rule goes wrong, namely, when the fleet is tight. In this case, $\nu>0$, and the optimal spot price rises above half the interactive price by $\frac12 g_m\nu/(\Delta\thS_m)$. Part (iv) shows that this happens for every model at once since the models share the fleet. For example, a launch-day surge on the large model raises the spot price of the small one, even though no small-model buyer has changed its behavior. Virtual machine types in a cloud spot market share a data center in the same way. One key difference is that the rate at which one virtual machine type displaces another is fixed by the machine size, whereas here it depends on the latency tier through $\rho_m$. Therefore, part (ii) also gives the provider a pricing floor for its interactive tier, i.e., $\pi_m\ge\rho_m\MC_m$.

\begin{proposition}[What a fixed discount loses]\label{prop:value}
Fix one model and one period, and let $c$ denote the marginal cost of a spot token, including any shadow price of capacity at the optimum. Let $p^\ast$ denote the optimal spot price and $\bar p$ any other price.
\begin{enumerate}[label=(\roman*)]
\item If capacity binds at neither price, the loss from charging $\bar p$ is $\frac{\Lambda}{\pi}(\bar p-p^\ast)^2$. Over periods $t=1,\dots,T$ in which this holds, a fixed price $\bar p$ loses $\sum_t\alpha_t(\bar p-p^\ast_t)^2\ \ge\ \bigl(\sum_t\alpha_t\bigr)\,\mathrm{Var}_\alpha(p^\ast)$, where $\alpha_t=\Lambda_t/\pi$ and $\mathrm{Var}_\alpha$ is the $\alpha$-weighted variance of the optimal price path.
\item If capacity $K$ binds at the optimum, so that $p^\ast$ is the market-clearing price, a lower price $\bar p<p^\ast$ with admissions rationed to $K$ loses $(p^\ast-\bar p)\,K$, i.e., the whole scarcity rent on the gap.
\end{enumerate}
\end{proposition}
\proofref{app:pf-value}

\begin{figure}[t]
\centering
\includegraphics[width=\textwidth]{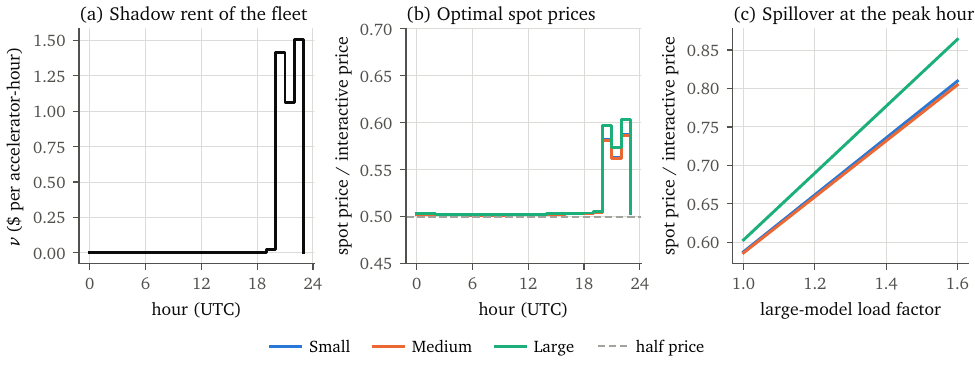}
\caption{One-period spot prices on a shared fleet (\cref{prop:static}), computed for each hour of a Monday at mean demand, with $D=1$ and no switching costs. (a) The shadow rent $\nu$ of the fleet is zero except in the $\stTightHours$ evening hours in which the fleet binds, and it peaks at \$$\stNuMax$ per accelerator-hour at $\stPeakHour$ UTC. (b) Optimal spot prices are half the interactive price when the fleet is slack, and they rise to $\stPmaxLarge$ of it for the large model at the peak. Over the day, a fixed half price gives up $\stLossPct\%$ of the optimal spot profit, and $\stLossTightPct\%$ of it in the tight hours. (c) Spillover. At the peak hour, scaling the interactive load of the large model alone raises the spot price of every model. At $\stSpillMult$ times its load, the spot prices of the small and medium models rise by $\stSpillSmall\%$ and $\stSpillMedium\%$.}
\label{fig:static}
\end{figure}

Therefore, the loss from a fixed discount is second order in slack periods and first order in tight periods. In other words, a half-price rule is close to optimal when there is spare capacity, whereas in tight periods it gives up the full scarcity rent while rate limits ration access. \Cref{fig:static} illustrates both regimes on the calibrated fleet. The same asymmetry underlies the classical case for peak-load pricing \citep{Steiner1957,Crew1995PeakLoad}, and it has been noted for flat-rate charging of cloud capacity, which is inefficient both when demand is far below capacity and when it exceeds capacity \citep{Zhang2011Spot}.

\subsection{Deferral: prices follow the cheapest capacity in the window}\label{sec:deferral}

Spot demand can be stored since work can wait until its deadline, whereas capacity cannot. To understand the effect of this asymmetry on prices, we fix the placement of one model so that spot capacities $K_t$ are given for $t=0,\dots,T-1$, and we assume that demand is known. The provider solves
\begin{equation}\label{eq:defer}
\max_{a,x\ge0}\ \sum_{t}\Bigl[a_t\pi\Bigl(1-\frac{a_t}{\Lambda_t}\Bigr)-c_tx_t-h\,q_{t+1}\Bigr]\quad\text{s.t. }x_t\le K_t\ [\mu_t],\ \eqref{eq:backlog}\text{ with }z\equiv0,\ \text{deadline }D,
\end{equation}
where $c_t$ is the energy cost of a spot token, $q_0=0$, and all work must be served by its deadline. Let $\mathcal W_t=\{t,\dots,\min(t+D,T)-1\}$ denote the service window of work admitted in period $t$.

\begin{theorem}[Deferral pricing]\label{thm:defer}
Problem \eqref{eq:defer} has an optimal solution with capacity multipliers $\mu_t\ge0$ such that, with the \emph{effective cost}
\begin{equation}\label{eq:Ct}
C_t=\min_{\tau\in\mathcal W_t}\bigl[c_\tau+\mu_\tau+h(\tau-t)\bigr],
\end{equation}
\begin{enumerate}[label=(\roman*)]
\item every period with $a_t>0$ has spot price $p_t=\frac12(\pi+C_t)$, and every period with $a_t=0$ has $C_t\ge\pi$;
\item work admitted in period $t$ can be scheduled so that it is served only in periods that attain the minimum in \eqref{eq:Ct};
\item if capacity is slack in some period $\tau\in\mathcal W_t$ ($\mu_\tau=0$), then $p_t\le\frac12\bigl(\pi+c_\tau+h(\tau-t)\bigr)$;
\item optimal profit is nondecreasing in $D$.
\end{enumerate}
\end{theorem}
\proofref{app:pf-defer}

\begin{figure}[t]
\centering
\includegraphics[width=\textwidth]{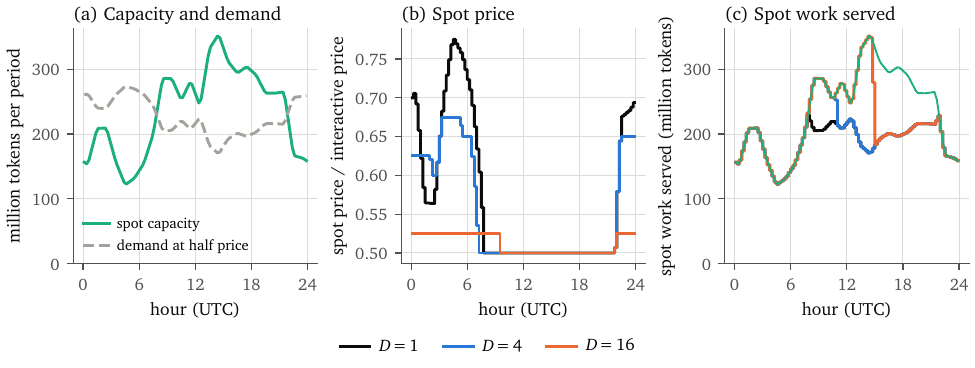}
\caption{Deferral pricing (\cref{thm:defer}) for the \dfModel{} model with its placement fixed at the best static level, on a day on which spot demand is 30\% above its mean. (a) Spot capacity left by interactive traffic, and spot demand at half price. The two move in opposite directions. (b) Optimal spot prices for deadlines of $D=1$, $4$, and $16$ periods (15 minutes, one hour, and four hours). The peak price falls from $\dfPmaxOne$ to $\dfPmaxFour$ and $\dfPmaxSixteen$ of the interactive price. In every period with sales, the computed price equals $\frac12(\pi+C_t)$ to within $\dfMaxDevTicks$ of a price tick $\pi/K$. (c) Spot work served. A longer deadline moves work out of the night-time trough of capacity into the following hours.}
\label{fig:defer}
\end{figure}

\Cref{thm:defer} shows that a buyer who accepts a $D$-period deadline buys the cheapest capacity in its window, valued at the shadow price of that later period plus the cost of waiting for it, and that the provider passes half of the saving on in the price. With $D=1$, $C_t=c_t+\mu_t$, and the spot price follows the current scarcity, as in \cref{prop:static}. With a longer deadline, a peak-hour price is capped by an off-peak cost (part iii, \cref{fig:defer}). In other words, deferral flattens spot prices and moves the scarcity rent from peak periods to the slack periods that absorb the deferred work. Hence, a deadline is a product attribute with a price. A buyer who is willing to wait buys capacity that would otherwise idle, and the provider shares the saving through a lower price for longer deadlines.

\subsection{Backfill: spot work makes reliability cheaper}\label{sec:backfill}

Placement is chosen before interactive load is known since weights take time to load. Therefore, a provider that wants interactive streams to meet their speed floor must hold safety capacity, which idles unless it is put to other use. Spot work provides such a use. Consider one model in one period with interactive load $R$ drawn from a continuous distribution $F$, $\bar F=1-F$. The provider chooses $n$ replicas at cost $k>0$ each (the fleet's shadow rent plus static energy), serves $\min(R,\thI n)$ interactive tokens per second, and sells the spot capacity left over by \cref{lem:cobatch}, which is worth $v\ge0$ per token at the margin. Its expected profit is given by
\begin{equation}\label{eq:newsvendor}
\Phi(n)=\Delta\Bigl[(\pi+\beta)\,\E\min(R,\thI n)-\beta\,\E R+v\,\thS\,\E\Bigl(n-\frac{R}{\thI}\Bigr)^{\!+}\Bigr]-k\,n .
\end{equation}

\begin{theorem}[Spot backfill raises the optimal service level]\label{thm:backfill}
Assume that $\rho v<\pi+\beta$ (an interactive token is worth more than the spot capacity it displaces) and $k<\Delta\thI(\pi+\beta)$. Then $\Phi$ is concave, and:
\begin{enumerate}[label=(\roman*)]
\item if $\Delta\thS v\ge k$, $\Phi$ is nondecreasing, i.e., spot work alone pays for the replica, and $n$ is set by the fleet;
\item otherwise, the optimal $n^\ast$ satisfies the critical-fractile condition
\begin{equation}\label{eq:fractile}
\Prob\bigl(R>\thI n^\ast\bigr)=\varpi(v):=\frac{k/(\Delta\thI)-\rho v}{(\pi+\beta)-\rho v};
\end{equation}
\item $\varpi(v)$ is strictly decreasing in $v$. The optimal probability of missing the interactive speed floor falls as spot capacity becomes more valuable. Without a spot market ($v=0$), it is the newsvendor fractile $k/(\Delta\thI(\pi+\beta))$;
\item if $R\sim N(\hat r,\sigma^2)$, then $n^\ast=(\hat r+z^\ast\sigma)/\thI$ with $z^\ast=\Phi_N^{-1}(1-\varpi(v))$, where $\Phi_N$ and $\phi_N$ are the standard normal distribution and density functions. The expected spot capacity is $\Delta\rho\,\sigma\,\bigl(z^\ast\Phi_N(z^\ast)+\phi_N(z^\ast)\bigr)$, which grows with the volatility $\sigma$ of interactive load.
\end{enumerate}
\end{theorem}
\proofref{app:pf-backfill}

\begin{figure}[t]
\centering
\includegraphics[width=\textwidth]{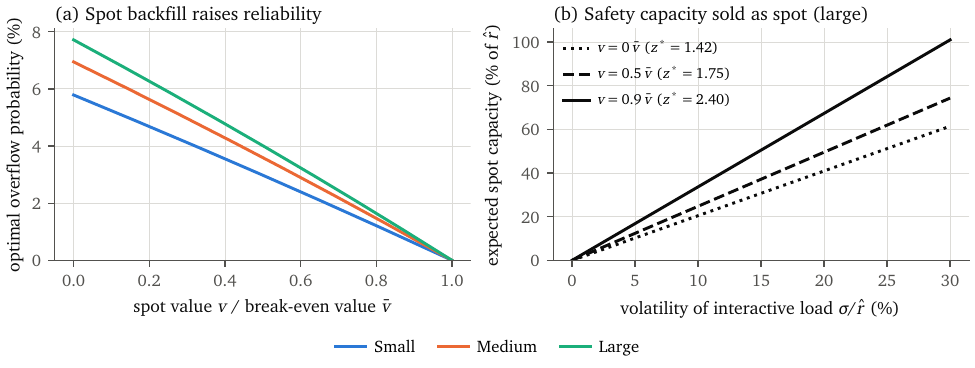}
\caption{Spot backfill (\cref{thm:backfill}) with replicas costed at \$$\accRent$ per accelerator-hour. (a) Optimal probability that interactive load overflows the placement, against the value $v$ of spot capacity relative to the break-even value $\bar v=k/(\Delta\thS)$ at which spot work alone pays for a replica. Without a spot market, the optimal overflow probabilities are $\bfZeroSmall\%$, $\bfZeroMedium\%$, and $\bfZeroLarge\%$; at $v=\bar v/2$, they fall to $\bfHalfSmall\%$, $\bfHalfMedium\%$, and $\bfHalfLarge\%$. (b) Expected spot capacity created by the safety margin of the large model, as a share of its mean interactive load, against the volatility of that load.}
\label{fig:backfill}
\end{figure}

The economics are those of the classic newsvendor \citep{Arrow1951}, but the salvage value of unused capacity is now provided by the spot market. When the spot market is liquid ($v$ high), safety capacity is cheap to hold since whatever the interactive tier does not use is sold. Therefore, a provider with a spot tier should promise, and deliver, better interactive service than one without (\cref{fig:backfill}a). Amazon's spot instances sell the unused part of a fleet sized for on-demand and reserved customers in the same way \citep{BenYehuda2013Spot}. One key difference for inference is that the unused capacity is mostly the safety margin of the interactive tier, and its size grows with the volatility of interactive demand (part iv). This is the mechanism by which the controller of \cref{sec:mpc} sizes placements, since its expected-overflow constraints embed \eqref{eq:fractile}.

\subsection{Placement: pricing transients and placing trends}\label{sec:hysteresis}

Moving accelerators between models is slow and costly, whereas changing a spot price is instant and free. To isolate this trade-off, we decompose \eqref{eq:P} by pricing the shared fleet. Given rents for accelerators and power, the value of holding $n$ replicas of model $m$ in period $t$ is a function $V_t(n)$, defined as the optimal value of that period's pricing and service problem net of the rents. This function is concave in $n$ since that problem is a concave program in which $n$ enters only through linear terms (\cref{lem:valueconcave} in Appendix~\ref{app:prelim}). The placement problem of one model is then given by
\begin{equation}\label{eq:placement}
\max_{n_0,\dots,n_{T-1}}\ \sum_{t=0}^{T-1}\Bigl[V_t(n_t)-\kappa^+(n_t-n_{t-1})^+-\kappa^-(n_{t-1}-n_t)^+\Bigr],\qquad n_{-1}\text{ given}.
\end{equation}

\begin{theorem}[Placement hysteresis]\label{thm:hyst}
Let each $V_t:\R\to\R$ be concave.
\begin{enumerate}[label=(\roman*)]
\item For any optimal plan $n^\ast$, there are supergradients $w_t\in\partial V_t(n^\ast_t)$ such that the \emph{continuation value} $\xi_t=\sum_{s=t}^{T-1}w_s$ satisfies $\xi_t\in[-\kappa^-,\kappa^+]$ for all $t$, with $\xi_t=\kappa^+$ whenever $n^\ast_t>n^\ast_{t-1}$ and $\xi_t=-\kappa^-$ whenever $n^\ast_t<n^\ast_{t-1}$.
\item Consequently, if a replica is loaded in period $t_1$ and the next change is an unload in period $t_2$, then $\sum_{s=t_1}^{t_2-1}w_s=\kappa^++\kappa^-$, i.e., the marginal replica earns exactly its round-trip switching cost over its tenure. Over any window $[t_1,t_2)$, $\bigl|\sum_{s=t_1}^{t_2-1}w_s\bigr|\le\kappa^++\kappa^-$.
\item \emph{(Spike absorption.)} Let $\bar n$ maximize a concave $V$, $n_{-1}=\bar n$, and $V_t=V$ except on a set $S$ of $d$ periods, on which the value of replicas beyond $\bar n$ rises by at most $\delta\ge0$ each: $V_t(n)-V_t(\bar n)\le V(n)-V(\bar n)+\delta\,(n-\bar n)^+$ for all $n$. If $d\,\delta\le\kappa^+$, the constant plan $n_t\equiv\bar n$ is optimal. If plans must end at $\bar n$, the same holds whenever $d\,\delta\le\kappa^++\kappa^-$.
\end{enumerate}
\end{theorem}
\proofref{app:pf-hyst}

\begin{figure}[t]
\centering
\includegraphics[width=\textwidth]{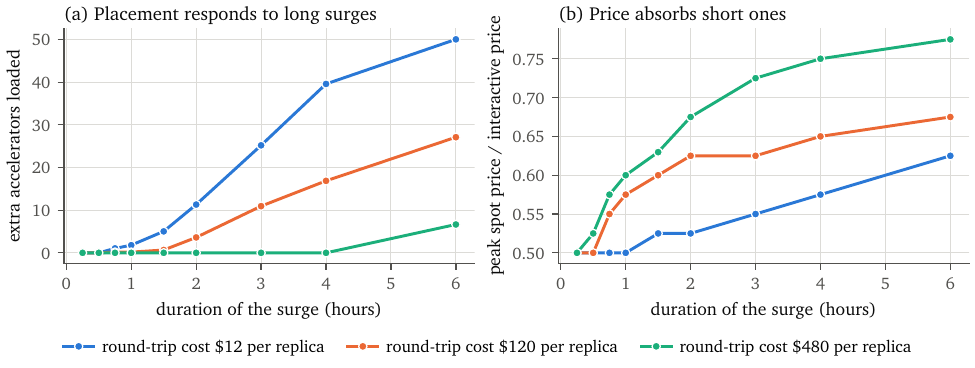}
\caption{Pricing of transients and placement of trends (\cref{thm:hyst}). Spot demand for the large model rises $\hySurge$-fold at $\hyStart$ UTC for a duration between 15 minutes and six hours, and the controller of \cref{sec:mpc} plans one day with perfect forecasts. (a) Accelerators loaded for the large model beyond the plan without the surge. (b) Peak spot price of the large model during the surge. With a round-trip switching cost of \$$\hyRTOne$, \$$\hyRTTen$, and \$$\hyRTForty$ per replica, the controller first moves at least half a replica for surges of $\hyFirstOne$, $\hyFirstTen$, and $\hyFirstForty$ hours, respectively. Shorter surges are absorbed by the price alone, which peaks at up to $\hyPmaxOne$, $\hyPmaxTen$, and $\hyPmaxForty$ of the interactive price.}
\label{fig:hyst}
\end{figure}

Part (iii) provides the operational rule, and \cref{fig:hyst} illustrates it. A transient that raises the marginal value of an extra replica by $\delta$ per period for $d$ periods is not worth a reload when $d\delta$ is below the round-trip cost. In this case, the provider absorbs the transient with the fast lever. The spot price rises (\cref{prop:static}(iv)), spot work is preempted and deferred within its deadline (\cref{thm:defer}), and interactive traffic keeps its speed. Only persistent shifts, such as a regional daily cycle or a launch that lasts for hours, justify moving weights. For large models, whose weights take minutes to load and whose replicas span many accelerators, $\kappa^\pm$ is large, and the band of inaction is wide. Note that modeling switching through explicit load and unload costs, rather than a quadratic penalty on capacity changes, yields the inaction bands familiar from investment under adjustment costs \citep{Dixit1989,Eberly1997}.

\begin{remark}[Price-taking providers]\label{rem:pricetaker}
Suppose that the provider sells spot capacity on a competitive exchange, where it can sell up to $\bar A_t$ tokens at a price $p^{\mathrm{ex}}_t$ that it does not control. The revenue term $a_t\pi(1-a_t/\Lambda_t)$ of \eqref{eq:defer} then becomes $p^{\mathrm{ex}}_ta_t$, with $a_t\le\bar A_t$. At an optimum, with $C_t$ defined by \eqref{eq:Ct} at the optimal multipliers, the provider sells the full depth $\bar A_t$ when $p^{\mathrm{ex}}_t>C_t$ and nothing when $p^{\mathrm{ex}}_t<C_t$. Therefore, it behaves as a seller whose ask equals its effective cost $C_t$, with no markup. The monopoly markup $\frac12(\pi-C_t)$ in \cref{prop:static} and \cref{thm:defer} is thus what market power adds to this ask.
\end{remark}
\proofref{app:pf-pricetaker}

%% file: sections/06_mpc.tex
\section{Model predictive control}\label{sec:mpc}

In this section, we develop the model predictive controller. The controller applies the receding-horizon principle of model predictive control \citep{Mayne2000,Rawlings2017}, which has been used to manage data centers and cloud capacity \citep{Kusic2009,Zhang2011Spot}. At the start of every period, the controller forecasts demand over the next $H$ periods, solves a deterministic program over that horizon, applies the first period's placement and prices, and repeats the procedure one period later with new information. Three features distinguish the program from those used for virtual machine capacity. First, the program has to handle a queue with deadlines. Second, the random interactive load enters through its expected overflow rather than its mean. Third, capacity changes carry explicit costs rather than a quadratic penalty. All three features can be expressed with linear constraints, and hence each period's problem is a linear program that solves in milliseconds.

\subsection{Forecasts}\label{sec:forecast}

For $k=0,\dots,H-1$ steps ahead, the controller requires forecasts $\hat r_{m,t+k}$ and $\sigma_{m,t+k}$ of the mean and standard deviation of interactive load, a forecast $\hat\Lambda_{m,t+k}$ of the spot intercept, and the electricity prices and power caps that the grid operator has announced. In the numerical study, interactive load is a seasonal profile $\bar R_{m,t}$ times a multiplicative deviation, i.e., $R_{m,t}=\bar R_{m,t}(1+\epsilon_{m,t})$, where $\epsilon$ is a first-order autoregression with coefficient $\phi$ and stationary standard deviation $\varsigma$. The controller estimates the current deviation from the last observed load and forecasts
\begin{equation}\label{eq:forecast}
\hat r_{m,t+k}=\bar R_{m,t+k}\bigl(1+\phi^{k+1}\hat\epsilon_{m,t-1}\bigr),\qquad
\sigma_{m,t+k}=\bar R_{m,t+k}\,\varsigma\sqrt{1-\phi^2}\,\Bigl(\sum_{i=0}^{k}\phi^{2i}\Bigr)^{1/2}.
\end{equation}
As a result, a flash crowd, which the seasonal profile does not contain, enters the forecast as a deviation that is expected to decay at rate $\phi$. The controller learns about a grid event when the operator announces it, which is $\evNotice$ hours ahead in the study. Note that any forecaster can replace \eqref{eq:forecast} since the program below needs only the means and standard deviations.

\subsection{The planning program}\label{sec:lp}

Fix the current period $t$ and index the horizon by $k=0,\dots,H-1$. We write $\hat r_{m,k}$ for $\hat r_{m,t+k}$, and similarly for the other forecasts. For every model and step, the decision variables are the replicas $n_{m,k}$, loads $u_{m,k}$, and unloads $v_{m,k}$; the expected interactive service $y_{m,k}$ and overflow $o_{m,k}$ in tokens per second; the spot admissions $a_{m,k}$, service $x_{m,k}$, and expiries $z_{m,k}$ in tokens; and a power shortfall $\omega_k$ that keeps the program feasible when an announced cap cannot be met. The effective replicas are $\tilde n_{m,k}=n_{m,k}-\ell_mu_{m,k}$ as in \eqref{eq:neff}.

\paragraph{Spot revenue.} Note that revenue $a\pi_m(1-a/\hat\Lambda)$ is concave in admissions. We split the admissions into $K$ segments of width $\hat\Lambda/K$, $a=\sum_{j=1}^Ka^{j}$ with $0\le a^{j}\le\hat\Lambda/K$, and credit segment $j$ with its average marginal revenue
\begin{equation}\label{eq:pwl}
\hat\Pi_K(a)=\max\Bigl\{\sum_{j=1}^K\pi_m\Bigl(1-\frac{2j-1}{K}\Bigr)a^{j}\ :\ \sum_ja^{j}=a,\ 0\le a^{j}\le\hat\Lambda/K\Bigr\}.
\end{equation}
Since the marginal revenues decrease in $j$, the linear program fills the segments in order, and no integer variables are needed.

\paragraph{Interactive service.} Placement is chosen before interactive load is known. The cost of too few replicas is captured by the expected overflow $\E(R-\thI\tilde n)^+$, which, for normal forecast errors, equals $\sigma L\bigl((\thI\tilde n-\hat r)/\sigma\bigr)$, where $L(z)=\phi_N(z)-z\bigl(1-\Phi_N(z)\bigr)$ is the standard normal loss function. Since $L$ is convex and decreasing, the expected overflow is bounded below by its tangents. With a grid $z_1<\dots<z_{\cutN}$ of $\cutN$ points from $-2.5$ to $3$, the program requires
\begin{equation}\label{eq:cuts}
o_{m,k}\ \ge\ \sigma_{m,k}L(z_i)-\bigl(1-\Phi_N(z_i)\bigr)\bigl(\thI_m\tilde n_{m,k}-\hat r_{m,k}-\sigma_{m,k}z_i\bigr),\qquad i=1,\dots,\cutN,
\end{equation}
together with $o_{m,k}\ge\hat r_{m,k}-y_{m,k}$, $0\le y_{m,k}\le\hat r_{m,k}$, and $y_{m,k}\le\thI_m\tilde n_{m,k}$. At the optimum, $y_{m,k}=\hat r_{m,k}-o_{m,k}$ is the expected interactive service $\E\min(R,\thI\tilde n)$, which is what the spot tier inherits.

\paragraph{The program.} With $\lambda^e_k=e_k\Delta/3600$ and the backlog $q_{m,k+1}=q_{m,0}+\sum_{j\le k}(a_{m,j}-x_{m,j}-z_{m,j})$, the controller solves
\begin{equation}\label{eq:lp}
\begin{aligned}
\max\ &\sum_{k=0}^{H-1}\Bigl\{\sum_m\Bigl[\hat\Pi_K(a_{m,k})-\Delta(\pi_m+\beta_m)\,o_{m,k}-(\pi_m+\gamma_m)\,z_{m,k}-h_mq_{m,k+1}\\
&\hspace{6.5em}-\kappa^+_mu_{m,k}-\kappa^-_mv_{m,k}\Bigr]-\lambda^e_k\,\Omega_k-c^\omega\omega_k\Bigr\}\\
\text{s.t. }\ & x_{m,k}\le\Delta\bigl(\thS_m\tilde n_{m,k}-\rho_m\,y_{m,k}\bigr)\quad[\mu_{m,k}],\qquad
\textstyle\sum_mg_mn_{m,k}\le G\quad[\nu_k],\qquad \Omega_k\le W_k+\omega_k\quad[\eta_k],\\
& u_{m,k}\ge n_{m,k}-n_{m,k-1},\qquad v_{m,k}\ge n_{m,k-1}-n_{m,k},\qquad\text{\eqref{eq:cuts} and the interactive bounds},\\
& \textstyle\sum_{j\le k}(x_{m,j}+z_{m,j})\ \ge\ d_{m,k}+\sum_{j\le k-D+1}a_{m,j},\qquad \sum_{j\le k}(x_{m,j}+z_{m,j})\ \le\ q_{m,0}+\sum_{j\le k}a_{m,j},\\
& q_{m,H}=0,\qquad \text{all variables}\ge0,
\end{aligned}
\end{equation}
where $\Omega_k$ is the power \eqref{eq:power} with $Y$ replaced by $y$, $n_{m,-1}$ is the current placement, $d_{m,k}$ is the part of the current backlog that falls due by step $k$, and $c^\omega$ is a large penalty for drawing more than the cap. The objective is the expected profit \eqref{eq:profit} with the exogenous interactive revenue $\Delta\pi_m\hat r_{m,k}$ dropped. In this objective, each token of overflow loses its price and incurs its penalty. Similarly, expired work loses its price, which is refunded, and incurs its penalty. The last three lines of constraints represent the deadlines, the availability of work, and a terminal condition that the horizon end with an empty backlog, which prevents the controller from pushing work beyond the horizon.

\begin{lemma}[Deadlines as cumulative constraints]\label{lem:fifo}
Fix one model, admissions $a_0,\dots,a_{H-1}$, and an initial backlog in which $d_k$ units fall due by step $k$. Serve the backlog in first-in-first-out order. A sequence of service and expiry $(x_k+z_k)_k$ is feasible and completes every unit by its deadline if and only if, for every $k$, cumulative service lies between the work due and the work available:
\[
d_k+\sum_{j\le k-D+1}a_j\ \le\ \sum_{j\le k}(x_j+z_j)\ \le\ q_0+\sum_{j\le k}a_j .
\]
\end{lemma}
\proofref{app:pf-fifo}

\Cref{lem:fifo} uses the fact that all spot work of a model has the same relative deadline so that first-in-first-out order is also earliest-deadline-first order. With heterogeneous deadlines, the same constraints hold for each deadline class, and the controller would need one queue per class.

\begin{lemma}[Approximation error]\label{lem:pwl}
\begin{enumerate}[label=(\roman*)]
\item $\hat\Pi_K$ is the piecewise-linear interpolation of $\Pi(a)=\pi a(1-a/\hat\Lambda)$ at the points $a=j\hat\Lambda/K$, and $0\le\Pi(a)-\hat\Pi_K(a)\le\pi\hat\Lambda/(4K^2)$ on $[0,\hat\Lambda]$. The error is at most $1/K^2$ of the largest revenue $\pi\hat\Lambda/4$, which is $\pwlErrPct\%$ for $K=\Kseg$.
\item The right-hand side of \eqref{eq:cuts}, maximized over $i$, is at most the expected overflow and equals it at the grid points. Between grid points $z_i<z_{i+1}$, the gap is at most $\sigma(z_{i+1}-z_i)^2\phi_N(0)/4$, and for the grid used it is at most $\cutErr\,\sigma$ on the grid's range.
\end{enumerate}
\end{lemma}
\proofref{app:pf-pwl}

Therefore, the program slightly understates spot revenue and slightly understates the risk of interactive overflow. Neither error matters at the resolution of the study. The planned price is obtained from the demand curve at the planned admissions, i.e., $p_{m,t}=\pi_m(1-a_{m,0}/\hat\Lambda_{m,0})$. Since the optimal admissions usually sit at a breakpoint of $\hat\Pi_K$, prices move in ticks of $\pi_m/K$, which is $2.5\%$ of the interactive price for $K=\Kseg$.

\subsection{The control loop}

\Cref{alg:mpc} summarizes the closed loop. Two steps convert the plan into actions. First, the planned replicas are rounded to whole numbers, rounding to the nearest integer where the fleet allows and down otherwise. Second, the planned price is posted, and buyers decide how much work to submit at that price. The program learns what buyers actually submitted only in the next period, when the realized admissions appear in the backlog. Therefore, any gap between planned and realized demand is absorbed by the queue and corrected by the next plan through feedback.

\begin{algorithm}[t]
\caption{Receding-horizon placement and pricing}\label{alg:mpc}
\begin{algorithmic}[1]
\Require horizon $H$, segments $K$, cut grid $z_1,\dots,z_{\cutN}$, current placement $n_{-1}$
\For{each period $t=0,1,\dots$}
  \State observe last period's interactive load and the backlog $q_{m,0}$ with its deadlines $d_{m,k}$
  \State update forecasts $\hat r,\sigma,\hat\Lambda$ by \eqref{eq:forecast}; read announced $e_k$ and $W_k$
  \State solve the linear program \eqref{eq:lp}; obtain $n_{m,0}$, $a_{m,0}$ and the multipliers $\mu,\nu,\eta$
  \State $n_{m,t}\gets$ whole-replica rounding of $n_{m,0}$; start loads and unloads
  \State post spot prices $p_{m,t}\gets\pi_m\bigl(1-a_{m,0}/\hat\Lambda_{m,0}\bigr)$
  \State during the period: serve interactive traffic first; serve the spot backlog first-in-first-out with the remaining capacity \eqref{eq:spotcap} and power; expire overdue work
\EndFor
\end{algorithmic}
\end{algorithm}

\paragraph{Relation to the structural results of \cref{sec:structure}.} The program \eqref{eq:lp} contains the four simplified problems of \cref{sec:structure} as special cases, and its optimality conditions reproduce their rules. For one step with $D=1$ and no switching costs, the program reduces to \eqref{eq:static}, up to the approximations of \cref{lem:pwl}, and hence its spot prices are $\frac12(\pi_m+\MC_m)$ with the shadow rent $\nu_k$ and power price $\eta_k$ of that step (\cref{prop:static}). With longer deadlines, the cumulative constraints of \cref{lem:fifo} allow work to move to the cheapest step in its window, and prices follow the effective cost $C_t$ of \cref{thm:defer}. The cuts \eqref{eq:cuts} set the interactive safety margin at the critical fractile of \cref{thm:backfill}, with the spot tier as salvage value. In particular, at the optimum, the slope $1-\Phi_N(z)$ of the active cut balances the cost of a replica against the value of its spot capacity. Moreover, the load and unload costs produce the inaction band of \cref{thm:hyst}. The multipliers $\mu_{m,k}$, $\nu_k$, and $\eta_k$ are also useful outputs in their own right. For example, $\rho_m\MC_m$ is a floor for interactive prices, and the effective cost $C_t$ is the provider's competitive ask on an exchange (\cref{rem:pricetaker}).

\paragraph{Size and speed.} With $M$ models and horizon $H$, the program has $(8+K)MH+H$ variables, which amounts to $\lpVars$ for $M=3$, $H=\Hbase$, and $K=\Kseg$, and about $30MH$ constraints, most of which are cuts. HiGHS \citep{Huangfu2018} solves it in about $\lpMs$ milliseconds on a laptop, compared with a control period of $15$ minutes. Fleets with tens of models and horizons of a day should remain well within the period since the program grows only linearly in both. In contrast, the cost of exact dynamic programming grows exponentially with the number of models and the length of the deadline.

%% file: sections/07_numerics.tex
\section{Numerical study}\label{sec:numerics}

In this section, we evaluate how much a provider gains from dynamic placement and pricing, where the gain comes from, and how it depends on forecasts, switching costs, and contract terms. All numbers come from the scripts listed in Appendix~\ref{app:repro}.

\subsection{Setup}

\paragraph{Fleet and models.} Consider a provider that runs $G=\fleetG$ accelerators under a nominal power cap of $\Wnom$ kW and serves three models (\cref{tab:params}). The small model fits on one accelerator, the medium model on two, and the large mixture-of-experts model on eight. Their batching parameters yield interactive batch caps between 83 and 133 sequences and exchange rates $\rho_m$ between $\rhoSmall$ and $\rhoMedium$. Loading weights takes $\loadSmall$, $\loadMedium$, and $\loadLarge$ seconds, which corresponds to $\ellSmall$, $\ellMedium$, and $\ellLarge$ of a 15-minute period. Each load costs \$$\kappaAcc$ and each unload \$$\kappaUnAcc$ per accelerator, which accounts for warm-up energy, drained sequences, and discarded caches. Interactive prices are \$$\priceSmall$, \$$\priceMedium$, and \$$\priceLarge$ per million tokens. An idle loaded accelerator draws $\wzero$ kW, a parked one draws $\wpark$ kW, and decode at full load adds $\wone$ kW.

\begin{table}[t]
\centering\small
\caption{Models in the numerical study. $g$: accelerators per replica. $\tau_0,\tau_1$: decode-step parameters of \eqref{eq:tau}, in milliseconds. $B$: memory cap on the batch. $\sI$: interactive speed floor (tokens per second per stream). $\bI$: interactive batch cap. $\thI,\thS$: replica throughput at full load in the interactive and spot modes, thousand tokens per second, with a spot speed floor of $\sS=\sSpot$. $\rho=\thS/\thI$. $L$: load time in seconds. $\pi$: interactive price, dollars per million tokens.}
\label{tab:params}
\begin{tabular}{@{}lrrrrrrrrrrr@{}}
\toprule
Model & $g$ & $\tau_0$ & $\tau_1$ & $B$ & $\sI$ & $\bI$ & $\thI$ & $\thS$ & $\rho$ & $L$ & $\pi$ \\
\midrule
\input{figures/param_rows.tex} 
\bottomrule
\end{tabular}
\end{table}

\paragraph{Demand.} Mean interactive load follows the hour-of-day profile of token volume on a public inference marketplace, computed from 30 days of hourly data, with a peak-to-trough ratio of $\hodPeakTrough$ \citep{Qubax2026API}. The medium model's profile is shifted by eight hours, as for a second region, and the large model's daily swing is half again as large. Weekend load is $\weekend$ of weekday load. At the mean, interactive traffic occupies $\occMean\%$ of the fleet on average. Note that real serving traces are bursty around their daily pattern \citep{Wang2024BurstGPT}. Accordingly, load deviates from its mean by a first-order autoregression with coefficient $\arPhi$ and standard deviation $\arSigma\%$, as in \eqref{eq:forecast}. Spot demand follows a flatter daily profile with an independent $\spotSigma\%$ noise every period, and it is scaled so that, at half the interactive price, it would occupy $\spotHalfShare\%$ of the fleet. Spot work must finish within $D=\deadlineD$ periods ($\deadlineMin$ minutes). A lost interactive token costs $\beta_m=\betaMult\,\pi_m$ in service credits and goodwill, and an expired spot token is refunded and costs $\gamma_m=\gammaMult\,\pi_m$.

\paragraph{Electricity and events.} Electricity costs \$$\eMean$ per kWh on average, with a daily swing of \$$\eAmp$ that peaks at 20:00 UTC. The simulated week contains two events that the seasonal profile does not anticipate. On Wednesday, a model launch doubles the large model's interactive load, ramping up from $\flashStart$ UTC over $\flashRamp$ hour, holding for $\flashHold$ hours, and decaying over $\flashDecay$ hours. On Friday, the grid operator calls a demand-response event. From $\evStart$ UTC for $\evHours$ hours, the facility must stay below $\evCapPct\%$ of its nominal cap ($\evCapKW$ kW), and electricity costs \$$\evPrice$ per kWh. The event is announced $\evNotice$ hours ahead.

\paragraph{Policies.} We compare six policies (\cref{tab:policies}). \emph{Posted} represents current practice, namely, the best static placement for a typical weekday and a spot tier sold at half the interactive price. It is a strong version of that practice since the same planning program sets a rate limit on spot admissions so that the posted tier never accepts more work than it expects to finish by the deadline. \emph{Dynamic price} and \emph{Dynamic placement} each add one of the two levers. \emph{Myopic} uses both levers with a persistence forecast, i.e., it expects current interactive load, spot demand, electricity price, and power cap to continue unchanged. \emph{MPC} is \cref{alg:mpc} with the forecasts \eqref{eq:forecast} and a horizon of $\Hbase$ periods ($\Hhours$ hours). \emph{Oracle} knows the realized future over twice that horizon and bounds what any controller could achieve with this program.

\begin{table}[t]
\centering\small
\caption{Policies. All policies use the same planning program \eqref{eq:lp} with different restrictions. ``Static'' placement refers to the best constant placement for two typical weekdays at mean demand.}
\label{tab:policies}
\begin{tabular}{@{}llll@{}}
\toprule
Policy & Spot price & Placement & Forecast of interactive load, demand, power \\
\midrule
Posted & half of $\pi_m$, rate-limited & static & seasonal and AR(1), for the rate limit \\
Dynamic price & dynamic & static & seasonal and AR(1) \\
Dynamic placement & half of $\pi_m$, rate-limited & dynamic & seasonal and AR(1) \\
Myopic & dynamic & dynamic & persistence \\
MPC & dynamic & dynamic & seasonal and AR(1); events once announced \\
Oracle & dynamic & dynamic & perfect, horizon $2\times\Hbase$ periods \\
\bottomrule
\end{tabular}
\end{table}

\paragraph{Simulation.} Each policy is run for one week at 15-minute periods on $\rSeeds$ random draws of interactive load and spot demand, with the same draws for every policy. The plant serves interactive traffic first at the effective replicas \eqref{eq:neff}, serves spot work first-in-first-out within the capacity \eqref{eq:spotcap}, and enforces the power cap by shedding spot work first and interactive work second. Profit is computed by \eqref{eq:profit} without the holding cost, which serves as a planning device. Gains are measured against Posted, paired by draw, and reported with standard errors across draws.

\subsection{Results}

\begin{table}[t]
\centering\small
\caption{One week of operation, means over $\rSeeds$ draws. Profit in thousands of dollars per week; gain over Posted in percent, with its standard error in parentheses; share of the Oracle's gain captured. SLO: share of interactive tokens served at the promised speed. Spot: tokens of spot work served, in trillions. Surplus: spot buyers' surplus, in thousands of dollars. Delay: mean and 95th percentile of the wait of a spot token, in minutes. Loads: accelerators loaded with new weights per week. Util.: share of accelerator time spent decoding at full-load equivalent.}
\label{tab:main}
\setlength{\tabcolsep}{4pt}
\begin{tabular}{@{}lrrrrrrcrr@{}}
\toprule
Policy & Profit & Gain (\%) & Share (\%) & SLO (\%) & Spot & Surplus & Delay & Loads & Util. (\%) \\
\midrule
\input{figures/main_rows.tex} 
\bottomrule
\end{tabular}
\end{table}

\paragraph{Profit gain of \rGainMPC\% from dynamic control.} \Cref{tab:main} summarizes the week. MPC earns \$$\rProfitMPC$ thousand, compared with \$$\rProfitPosted$ thousand for Posted, which is a gain of $\rGainMPC\%$ (standard error $\rGainSeMPC$), and it captures $\rShareMPC\%$ of the gain of the Oracle. Of the \$$\rDecTotal$ thousand gained per week, \$$\rDecPenI$ thousand comes from interactive penalties avoided, \$$\rDecRevI$ thousand from interactive revenue no longer lost, and \$$\rDecRevS$ thousand from additional spot revenue, while switching costs take back \$$\rDecSwap$ thousand, and energy, refunds, and expiry penalties take back \$$\rDecOther$ thousand. In addition, the interactive tier receives better service. MPC misses the speed promise on $\rMissMPC\%$ of interactive tokens, compared with $\rMissPosted\%$ for Posted. Note that the two levers are complements. Dynamic prices alone gain $\rGainDynamicprice\%$, dynamic placement alone $\rGainDynamicplacement\%$, and both together under the same forecasts $\rGainMPC\%$. Forecasts of the daily cycle add a little. The Myopic controller gains $\rGainMyopic\%$, with more than half again as many accelerator loads as MPC, since it keeps reacting to the load that it has just seen.

\begin{figure}[t]
\centering
\includegraphics[width=\textwidth]{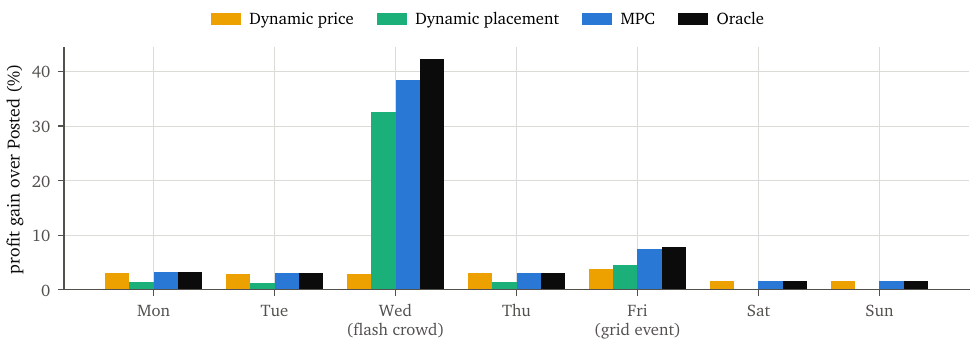}
\caption{Profit gain over Posted by day, means over $\rSeeds$ draws. The large model's flash crowd falls on Wednesday, and the grid event falls on Friday.}
\label{fig:days}
\end{figure}

\begin{figure}[p]
\centering
\includegraphics[width=\textwidth]{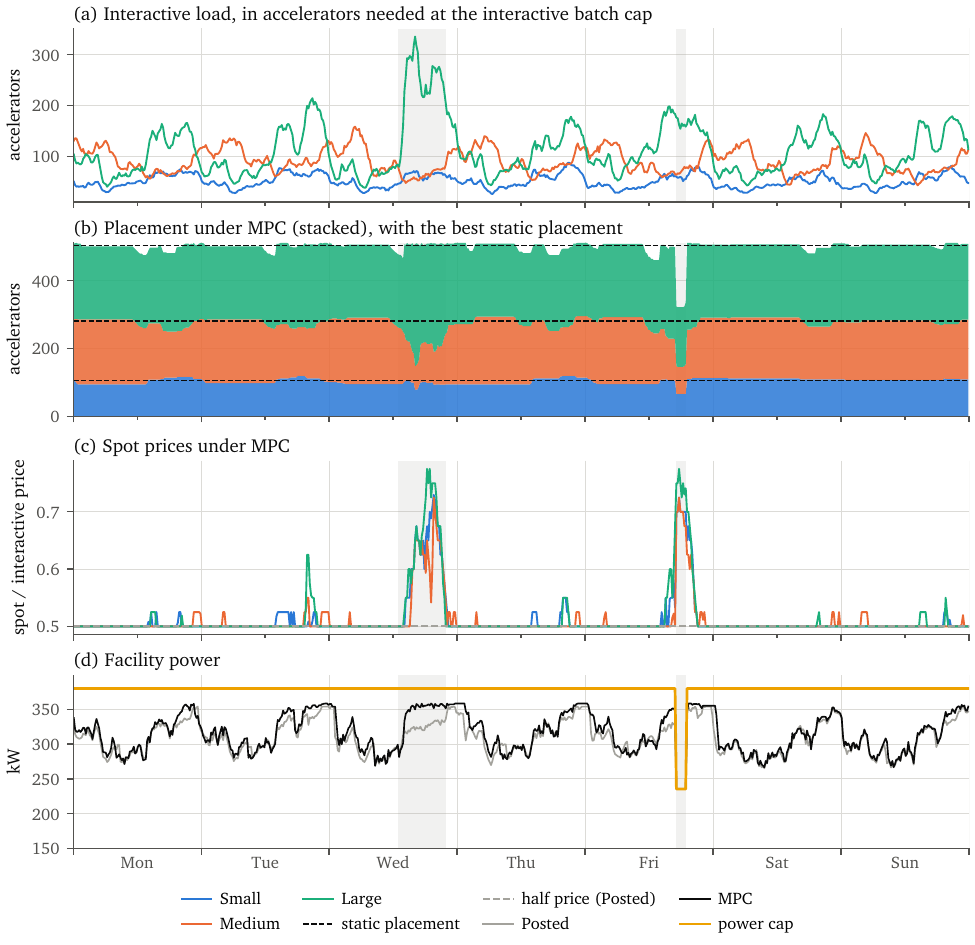}
\caption{One week under MPC (first draw). Gray bands mark the flash crowd on Wednesday and the grid event on Friday. (a) Interactive load of each model, in accelerators needed at the interactive batch cap. (b) Accelerators loaded with each model, stacked, where dashed lines mark the best static placement used by Posted. (c) Spot prices relative to the interactive price. (d) Facility power under Posted and MPC, and the power cap.}
\label{fig:week}
\end{figure}

\paragraph{Prices on ordinary days and placement during events.} \Cref{fig:days} breaks the gain down by day. On the five days without events, MPC gains $\rDayNormalMPC\%$, and dynamic prices alone already gain $\rDayNormalDynamicprice\%$. On these days, the fleet is tight only for a few evening hours. As \cref{prop:value} predicts, the fixed discount loses little when capacity is slack, and its losses are concentrated in the tight hours. On the day of the flash crowd, MPC gains $\rDayFlashMPC\%$, and on the day of the grid event, it gains $\rDayEventMPC\%$. Prices cannot help on these days since the interactive tier is the one short of capacity, and only moving accelerators or power to it can help. \Cref{fig:week} shows the week under MPC. Placement follows the daily cycle in small steps and moves sharply only for the two events. Spot prices stay at or within a tick or two of half price for most of the week and rise during the events.

\begin{figure}[t]
\centering
\includegraphics[width=\textwidth]{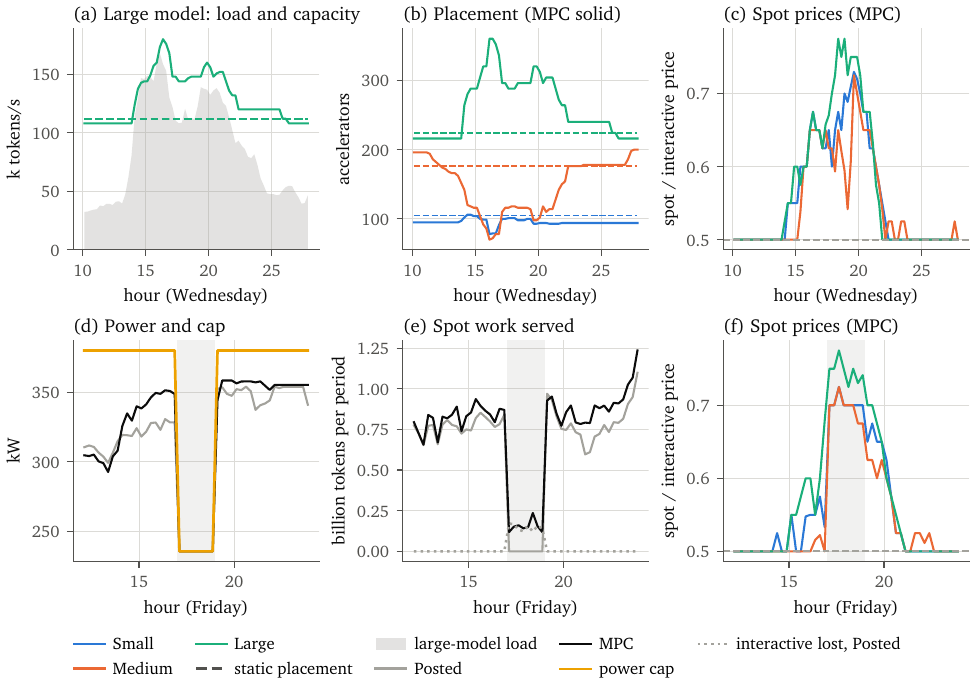}
\caption{The two events (first draw). Top: the flash crowd. (a) Large-model interactive load and the interactive capacity of its replicas under MPC (solid) and the static placement (dashed). (b) Accelerators per model under MPC (solid) and static (dashed). (c) Spot prices under MPC. Bottom: the grid event (gray band). (d) Facility power and the cap. (e) Spot work served under MPC and Posted, and interactive work that Posted cannot serve within the cap. (f) Spot prices under MPC.}
\label{fig:events}
\end{figure}

\paragraph{The flash crowd.} When the large model's load doubles, MPC moves accelerators from the medium model, whose load is low at that hour, to the large one, namely, from $\rFlashAccBefore$ to $\rFlashAccPeak$ accelerators at the peak, compared with a static allocation of $\rFlashStatic$ (\cref{fig:events}a,b). Since the forecast learns the surge from the load it observes, the first periods are served from the safety margin. In the first draw, Posted fails to serve $\rFlashLostPosted$ billion interactive tokens during the surge, and MPC fails to serve $\rFlashLostMPC$ billion. Dynamic prices alone fail to serve $\rFlashLostDynPrice$ billion, the same as Posted. The reason is that raising the spot price frees spot capacity on the large model's replicas, but the interactive tier needs more replicas rather than more room on the ones it has. At the same time, MPC raises the spot price of every model, to $\rFlashPmaxLarge$ of the interactive price for the large model and to $\rFlashPmaxSmall$ and $\rFlashPmaxMedium$ for the small and medium ones, respectively (\cref{fig:events}c). The other models' prices rise because they gave up accelerators, which is the spillover of \cref{prop:static}(iv).

\paragraph{The grid event.} Two hours before the event, MPC learns that the cap will fall to $\evCapKW$ kW. It raises spot prices to slow admissions, serves more of the backlog before the event, and, at the start of the event, unloads replicas to park their accelerators, going from $\rEvAccPre$ to $\rEvAccMPC$ loaded accelerators (\cref{fig:week}b). Facility power falls by $\rEvShedKW$ kW from its level in the two hours before (\cref{fig:events}d). The interactive tier is served in full throughout, and the spot tier absorbs the curtailment. In particular, MPC serves $\rEvSpotMPC$ billion spot tokens in the two hours of the event, compared with $\rEvSpotPreMPC$ billion in the two hours before. In contrast, Posted keeps its replicas loaded, cannot fit even its interactive load under the cap, and loses $\rEvLostPosted$ billion interactive tokens in the first draw ($\rEvShedPostedMean$ billion on average). The spot price during the event reaches $\rEvPmaxLarge$ of the interactive price for the large model (\cref{fig:events}f). In this sense, the spot tier serves as the facility's flexible load, and its price tells buyers when the facility is curtailed.

\paragraph{Buyers.} Spot buyers are not worse off. Although dynamic prices are above half price in tight hours, MPC serves more spot work in total. The posted tier rations by a rate limit that is set before demand is known, and it turns work away whenever demand exceeds the limit. In contrast, a price admits whatever buyers want at that price and lets the backlog absorb the surprise. Moreover, MPC has more capacity when it is needed since it moves replicas. It serves $\rSpotTokMPC$ billion spot tokens a week, compared with $\rSpotTokPosted$ billion for Posted, and spot buyers' surplus is $\rSurplusGain\%$ higher. Profit plus buyer surplus rises by $\rWelfareGain\%$. The cost to buyers is a longer wait. The mean delay of a spot token rises from $\rDelayPosted$ to $\rDelayMPC$ minutes, and its 95th percentile rises from $\rPninefivePosted$ to $\rPninefiveMPC$ minutes, which is still well within the $\deadlineMin$-minute deadline. Only $\rExpiredMPC\%$ of admitted spot work expires. The Oracle waits longest since it knows best when capacity will be free.

\subsection{Sensitivity}\label{sec:sens}

\begin{figure}[t]
\centering
\includegraphics[width=\textwidth]{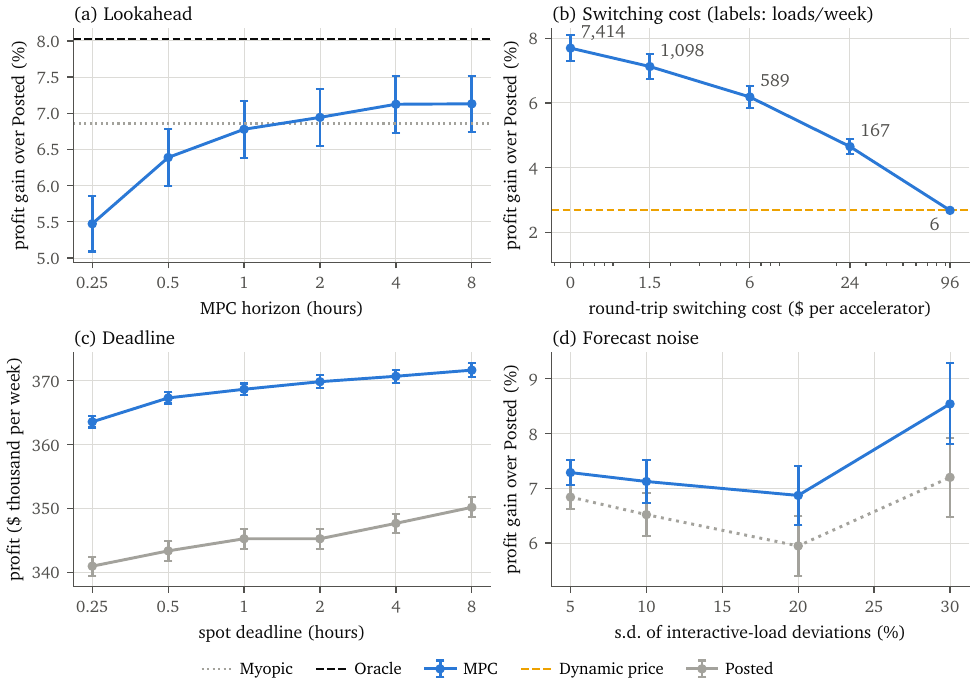}
\caption{Sensitivity, 10 draws per point, bars of $\pm$ one standard error. (a) Gain of MPC over Posted against the horizon of the program, with Myopic and Oracle at the base case. (b) Gain of MPC against the round-trip cost of loading and unloading one accelerator (the base case is \$1.50; zero is plotted at the left end); labels give accelerator loads per week. (c) Weekly profit of MPC and Posted against the spot deadline. (d) Gains of MPC and Myopic against the standard deviation of interactive-load deviations.}
\label{fig:sens}
\end{figure}

In this subsection, we vary one feature at a time on 10 draws, with gains paired by draw against Posted under the same conditions (\cref{fig:sens}).

\paragraph{Lookahead.} A controller that plans one period ahead gains $\rHOne\%$. Planning one hour ahead gains $\rHFour\%$, and planning four hours ahead gains $\rHSixteen\%$, with no further gain from eight hours (\cref{fig:sens}a). A few hours suffice since the deadline is two hours and placement transients last a few hours.

\paragraph{Switching costs and loading time.} Without switching costs, MPC gains $\rSwZero\%$ but loads $\rSwLoadsZero$ accelerators a week. The base cost cuts loads to $\rSwLoadsOne$ for a gain of $\rSwOne\%$. As the cost rises, placement freezes (\cref{thm:hyst}). At 64 times the base cost, the controller loads $\rSwLoadsSixtyFour$ accelerators a week and gains $\rSwSixtyFour\%$, which is the gain of dynamic prices alone (\cref{fig:sens}b). Loading time matters much less than loading cost. Scaling all load times by 0, 1, 2, and 3 yields gains of $\rLtZero\%$, $\rLtOne\%$, $\rLtTwo\%$, and $\rLtThree\%$, since the controller loads ahead of need, and the lost fraction of a period matters only when a load is a surprise.

\paragraph{Deadlines.} Longer deadlines raise the profit of both policies (\cref{fig:sens}c), as predicted by \cref{thm:defer}(iv). With a deadline of one period, MPC earns \$$\rDOneMPC$ thousand and Posted \$$\rDOnePosted$ thousand; with $\rDMaxHours$ hours, they earn \$$\rDMaxMPC$ thousand and \$$\rDMaxPosted$ thousand, respectively. The gain of MPC over Posted is $\rDGainOne\%$ and $\rDGainMax\%$ at the two ends. Deferral and dynamic pricing are partial substitutes since a long deadline allows even a fixed-price tier to move work out of tight hours. This observation is consistent with current batch tiers, which pair a fixed discount with a 24-hour window.

\paragraph{Forecast noise.} With a standard deviation of interactive-load deviations of $\rNoiseLowPct\%$, MPC gains $\rNoiseLowMPC\%$ and Myopic $\rNoiseLowMyopic\%$. At $\rNoiseHighPct\%$, they gain $\rNoiseHighMPC\%$ and $\rNoiseHighMyopic\%$, respectively (\cref{fig:sens}d). The gain does not shrink with noise since noise hurts the static policy more. A static placement must cover the unconditional variation of load, whereas the controller re-plans every period and sizes its margin to the error of a forecast a few periods ahead, which is much smaller when deviations persist. Both sell their margins as spot capacity, as in \cref{thm:backfill}.

%% file: figures/param_rows.tex
Small & 1 & 4 & 0.045 & 256 & 100 & 133 & 13.33 & 16.49 & 1.24 & 30 & 0.15 \\
Medium & 2 & 10 & 0.120 & 256 & 50 & 83 & 4.17 & 6.29 & 1.51 & 90 & 0.80 \\
Large & 8 & 20 & 0.100 & 256 & 30 & 133 & 4.00 & 5.61 & 1.40 & 300 & 3.00 \\

%% file: figures/main_rows.tex
Posted & 344.1 & --- & --- & 99.75 & 0.50 & 54.7 & 11 / 33 & 0 & 79.3 \\
Dynamic price & 353.4 & +2.7 (0.04) & 34 & 99.75 & 0.52 & 56.0 & 16 / 56 & 0 & 81.0 \\
Dynamic placement & 361.9 & +5.2 (0.30) & 64 & 99.98 & 0.51 & 57.0 & 10 / 26 & 1,238 & 81.2 \\
Myopic & 367.8 & +6.9 (0.32) & 85 & 99.99 & 0.51 & 55.8 & 9 / 22 & 1,688 & 81.7 \\
MPC & 369.8 & +7.4 (0.32) & 93 & 99.99 & 0.53 & 58.3 & 15 / 51 & 1,078 & 82.6 \\
Oracle & 371.8 & +8.0 (0.35) & 100 & 100.00 & 0.53 & 59.3 & 21 / 95 & 796 & 83.0 \\

%% file: sections/08_discussion.tex
\section{Discussion}\label{sec:discussion}

In this section, we discuss the implications of our results, their implementation, and their limitations.

\paragraph{Implications for providers.} Four operating rules follow from \cref{sec:structure}, and each of them continues to hold in the closed loop of \cref{sec:numerics}.
\begin{enumerate}[leftmargin=*,label=\arabic*.]
\item \emph{Price spot work midway between the interactive price and its effective cost.} The effective cost $C_t$ is the cheapest cost of serving the work anywhere in its deadline window, including the shadow rent of the fleet and of power in that period (\cref{thm:defer}). When the fleet is slack, the resulting spot price is close to the prevailing half price (\cref{prop:static}(v)). When the fleet is tight, the price is higher, and it is higher for every model at once since the models share the fleet.
\item \emph{Price interactive service above $\rho_m\MC_m$.} An interactive token displaces $\rho_m$ spot tokens (\cref{lem:cobatch}), and hence the shadow cost of interactive capacity is the spot cost scaled by the latency exchange rate (\cref{prop:static}(ii)). A provider that prices its fast tiers off its standard tier by a fixed multiple ignores the scarcity of the fleet.
\item \emph{Hold more safety capacity when a spot tier can use it.} Spot work serves as the salvage market for the interactive tier's safety margin, and hence the optimal probability of missing the speed promise falls as the spot tier becomes more valuable (\cref{thm:backfill}). In other words, the spot tier and the interactive tier's reliability are complements rather than rivals.
\item \emph{Price the transients and place the trends.} Move weights only when the cumulative value of a replica over the expected duration of a shift exceeds the round-trip cost of loading and unloading it (\cref{thm:hyst}). Absorb shorter shifts with the spot price and the backlog.
\end{enumerate}

\paragraph{Implementation.} The controller requires forecasts of means and standard deviations, the current backlog with its deadlines, and the announced electricity prices and power caps. Its program is a linear program of a few thousand variables that solves in milliseconds, and hence it can run every few minutes alongside an existing batch scheduler. Note that the scheduler needs no change. It serves interactive traffic first and fills the remaining slots with spot work in first-in-first-out order, which is what continuous-batching servers already do with priorities. Moreover, the spot price can be published as a schedule over the next few hours, which informs buyers of when capacity is likely to be cheap and when the facility expects to curtail.

\paragraph{Market power and the half-price rule.} Our provider is a monopolist of its own spare capacity facing linear demand, and the factor $\frac12$ in the pricing rules comes from that demand curve. With other valuation distributions, the markup over $C_t$ changes, but the cost side of every result does not. In particular, the shadow prices of the fleet and power, the effective cost under deferral, the critical fractile, and the inaction band are properties of the provider's operations rather than of its demand. If the provider sold its spare capacity on a competitive exchange, it would ask $C_t$ itself (\cref{rem:pricetaker}), and the difference between the monopoly price and $C_t$ measures the provider's market power over its spot tier.

\paragraph{The spot tier as flexible load.} Grid operators increasingly ask large new loads to curtail during the few hours a year when the grid is stressed \citep{Norris2025}. The grid event of \cref{sec:numerics} shows that a spot tier makes such curtailment cheap for an inference provider. The interactive tier keeps its promise, and the curtailment falls on work whose buyers accepted a deadline and are compensated for it through a lower price. The shadow price $\eta$ of power in the planning program represents the provider's value of a kilowatt at each hour. A provider could use it to bid its flexibility into a demand-response program, which would turn the spot tier's flexibility into revenue from the grid as well as from buyers.

\paragraph{Limitations.} The model is a fluid model. It summarizes batching by full-load throughputs and does not track the latency of individual requests, the split between prefill and decode, or the memory of key--value caches beyond the batch cap. It treats interactive demand as exogenous in the short run and spot buyers as price-takers who do not time their submissions against the price schedule. Moreover, forecast errors are normal, and there is one facility with one kind of accelerator. The planning program is solved by receding horizon without closed-loop guarantees. In addition, the numerical study is stylized. Its daily profile is measured, but its fleet, models, prices, and events are illustrative, and hence its percentages indicate magnitudes rather than predict a provider's results. Each of these limitations points to a direction for further work. Possible directions include several facilities with transfer between them, as in \citet{Zhang2013Placement}; prefill and decode as two resources; demand curves learned online; buyers who choose deadlines from a menu of prices; and robust or stochastic versions of the program.

%% file: sections/09_conclusion.tex
\section{Conclusion}\label{sec:conclusion}

Inference providers operate fleets sized for the interactive peak and sell the remaining capacity through fixed discounts. In this paper, we have developed a framework for allocating and pricing this capacity dynamically. Four features distinguish inference from conventional cloud capacity: capacity depends on the latency tier, placement is slow and costly while the split between tiers is free, demand can wait until a deadline, and power can bind. We have shown that, when these features are taken into account, the provider's problem has a clear structure. Spot prices lie midway between the interactive price and the effective cost of capacity in the deadline window. Interactive capacity carries a shadow cost scaled by the latency exchange rate. A spot tier raises the optimal reliability of the interactive tier. Placement should follow persistent shifts in demand and leave transients to prices. Building on these elements, we have designed a model predictive controller that solves a small linear program each period. In a simulated week, the controller raises profit by $\rGainMPC\%$ over a fixed half-price policy, captures $\rShareMPC\%$ of the gain available with perfect forecasts, keeps the interactive promise through a flash crowd and a grid curtailment, and leaves spot buyers better off.

More broadly, the analysis suggests that the price of spare inference capacity should be understood as an operational signal rather than a fixed discount. This price summarizes, at each moment, the scarcity of accelerators, the cost and availability of power, and the flexibility that buyers offer through their deadlines. As inference becomes a basic input to economic activity and a significant load on the power grid, mechanisms that make this signal explicit will matter for the efficiency of AI services and for their integration with the energy system. As future work, we would extend the framework to facilities linked by transfers of work, to prefill and decode as separate resources, and to demand curves learned online. Other directions include buyers who choose deadlines strategically from a menu of prices and competition among providers whose spot tiers serve overlapping buyers.

%% file: sections/appendix.tex
\section{Proofs}\label{app:proofs}

In this appendix, we prove the results of \cref{sec:model,sec:structure,sec:mpc}. \Cref{app:prelim} collects the facts that the proofs use. The proofs then follow in the order in which the results appear, except that \cref{lem:fifo} is proved before \cref{thm:defer}, which uses it, and \cref{rem:pricetaker} is proved after \cref{thm:hyst}.

\subsection{Preliminaries}\label{app:prelim}

\paragraph{Supergradients and optimality conditions.} For a concave function $V$ on $\R$, $\partial V(n)=\{w:\ V(m)\le V(n)+w(m-n)\ \forall m\}$ denotes its superdifferential, which is the interval between the right and left derivatives of $V$ at $n$. Note that all optimization problems below maximize a concave objective subject to linear constraints. For such problems, the Karush--Kuhn--Tucker (KKT) conditions, with supergradients in place of gradients where the objective is not differentiable, are necessary at an optimum without any further constraint qualification, and they are sufficient for optimality \citep[Sections~28--29]{Rockafellar1970}. We use this fact without further mention.

\paragraph{Decode throughput.} With $\tau_m(b)=\tau_{0,m}+\tau_{1,m}b$ and $\theta_m(b)=b/\tau_m(b)$,
\[
\theta_m'(b)=\frac{\tau_{0,m}}{\tau_m(b)^2}>0,\qquad \theta_m''(b)=-\frac{2\tau_{0,m}\tau_{1,m}}{\tau_m(b)^3}\le0,
\]
and hence throughput is increasing and concave in the batch size, as stated in \cref{sec:fleet}. The speed of each stream, $1/\tau_m(b)$, is decreasing in $b$, and therefore a speed floor $s$ is met exactly by the batches $b\le(1/s-\tau_{0,m})/\tau_{1,m}$. Together with the memory cap $B_m$, this condition yields the batch caps $\bI_m$ and $\bS_m$ of \cref{sec:fleet}.

\paragraph{Concave value functions.} The placement problem of \cref{sec:hysteresis} relies on the following standard fact.

\begin{applemma}[Concave value of a parametric concave program]\label{lem:valueconcave}
Let $C\subseteq\R^p\times\R^q$ be convex and $f:C\to\R$ concave. For $n$ in the projection $N$ of $C$ onto $\R^q$, let $V(n)=\sup\{f(x,n):(x,n)\in C\}$. If $V$ is finite on $N$, then $V$ is concave on $N$.
\end{applemma}

\emph{Proof.} Let $n,n'\in N$, $\lambda\in[0,1]$ and $\varepsilon>0$, and choose $x,x'$ with $(x,n),(x',n')\in C$, $f(x,n)\ge V(n)-\varepsilon$ and $f(x',n')\ge V(n')-\varepsilon$. By convexity of $C$, the point $(\lambda x+(1-\lambda)x',\lambda n+(1-\lambda)n')$ lies in $C$, and by concavity of $f$,
\[
V\bigl(\lambda n+(1-\lambda)n'\bigr)\ \ge\ f\bigl(\lambda x+(1-\lambda)x',\lambda n+(1-\lambda)n'\bigr)\ \ge\ \lambda V(n)+(1-\lambda)V(n')-\varepsilon .
\]
Since $\varepsilon$ is arbitrary, $V$ is concave. \qed

In \cref{sec:hysteresis}, $V_t(n)$ is defined as the optimal value of one period's pricing and service problem with $n$ replicas, net of the rents. That problem maximizes an objective that is jointly concave in its decisions and in $n$ (a concave quadratic revenue less linear costs) over a polyhedron in which $n$ enters linearly. Therefore, \cref{lem:valueconcave} shows that $V_t$ is concave, as assumed in \cref{thm:hyst}.

\subsection{Proof of \texorpdfstring{\protect\cref{lem:cobatch}}{Lemma 1} (co-batching)}\label{app:pf-cobatch}

Consider any way of serving the load with $n$ divisible replicas, or equivalently with replicas that divide their time between batch compositions. Let $n_I$ denote the replica capacity that holds at least one interactive stream. Since every stream in a batch advances at $1/\tau_m(b)$ tokens per second, a replica that holds an interactive stream must run batches with $1/\tau_m(b)\ge\sI_m$, that is, $b\le\bI_m$. Its total throughput is then at most $\theta_m(\bI_m)=\thI_m$ since $\theta_m$ is increasing. Therefore, interactive load $r$ requires $n_I\ge r/\thI_m$, and those replicas have at most $\thI_mn_I-r$ tokens per second left for spot streams. The other $n-n_I$ replicas run spot work only, at most at batch $\bS_m$ and throughput $\thS_m$. Hence, total spot throughput is at most
\[
\thI_mn_I-r+\thS_m(n-n_I)=\thS_mn-r-(\thS_m-\thI_m)\,n_I\ \le\ \thS_mn-r-(\thS_m-\thI_m)\frac{r}{\thI_m}=\thS_mn-\rho_mr,
\]
using $\thS_m\ge\thI_m$ and $n_I\ge r/\thI_m$. The bound is attained with $n_I=r/\thI_m$ replicas at batch $\bI_m$ carrying interactive streams only and the rest at batch $\bS_m$ carrying spot streams only.

With whole replicas, let $k=\lceil r/\thI_m\rceil\le n$ replicas run at batch $\bI_m$ and carry all interactive streams, with their spare slots filled by spot streams, and let the other $n-k$ replicas run spot work at batch $\bS_m$. Spot throughput is $k\thI_m-r+(n-k)\thS_m=\thS_mn-r-(\thS_m-\thI_m)k$. Since $k<r/\thI_m+1$, this exceeds $\thS_mn-\rho_mr-(\thS_m-\thI_m)$.

Finally, if neither batch cap is set by memory, $\tau_m(\bI_m)=1/\sI_m$, and hence $\thI_m=\bI_m\sI_m=(1-\tau_{0,m}\sI_m)/\tau_{1,m}$, and likewise $\thS_m=(1-\tau_{0,m}\sS)/\tau_{1,m}$. Their ratio is $\rho_m=(1-\tau_{0,m}\sS)/(1-\tau_{0,m}\sI_m)$. \qed

\subsection{Proof of \texorpdfstring{\protect\cref{prop:static}}{Proposition 1} (spot prices on a shared fleet)}\label{app:pf-static}

\emph{Existence.} The capacity constraints imply $n_m\ge\rho_mR_m/\thS_m=R_m/\thI_m$, and the assumptions make the point $n_m=R_m/\thI_m$, $a_m=0$ feasible. The feasible set is closed, and it is bounded since $n$ is bounded by the fleet and $a$ by capacity. Since the objective is continuous, an optimum exists.

\emph{KKT conditions.} We attach multipliers $\mu_m\ge0$ to the capacity constraints, $\nu\ge0$ to the fleet, and $\eta\ge0$ to power. Let $\epsilon_m=g_m(\lambda^e+\eta)w^1/(\Delta\thS_m)$ denote the energy and power cost of the decode work of one spot token. Stationarity in $n_m\ge0$ and $a_m\ge0$ reads
\begin{align}
\mu_m\Delta\thS_m-g_m\bigl[\nu+(\lambda^e+\eta)(w^0-w^{\mathrm{park}})\bigr]&\le0,\quad\text{with equality if }n_m>0, \label{eq:kkt-n}\\
\pi_m(1-2a_m/\Lambda_m)-\epsilon_m-\mu_m&\le0,\quad\text{with equality if }a_m>0. \label{eq:kkt-a}
\end{align}

\emph{(i)} Suppose that $a_m>0$. Then, $n_m>0$, and hence \eqref{eq:kkt-n} holds with equality, and $\mu_m+\epsilon_m=\MC_m(\nu,\eta)$ by the definition \eqref{eq:MC} and $\wfull=w^0-w^{\mathrm{park}}+w^1$. Then, \eqref{eq:kkt-a} gives $\pi_m(1-2a_m/\Lambda_m)=\MC_m$, and therefore $\MC_m<\pi_m$, $a_m=\frac12\Lambda_m(1-\MC_m/\pi_m)$, and, from \eqref{eq:demand}, $p_m=\pi_m(1-a_m/\Lambda_m)=\frac12(\pi_m+\MC_m)$. Conversely, suppose that $a_m=0$. By \eqref{eq:kkt-n}, $\mu_m+\epsilon_m\le\MC_m$ whether or not $n_m>0$, and by \eqref{eq:kkt-a}, $\pi_m\le\mu_m+\epsilon_m$. Hence, $\MC_m\ge\pi_m$. Therefore, model $m$ sells spot work if and only if $\MC_m<\pi_m$.

\emph{(ii)} Let $\Pi(R)$ denote the optimal profit as a function of the interactive loads, and let $\mathcal L(x,\lambda;R)$ denote the Lagrangian of \eqref{eq:static} with decision $x=(n,a)$ and multipliers $\lambda=(\mu,\nu,\eta)$. The load $R_m$ enters $\mathcal L$ linearly and separately from $x$, through the energy term, the capacity constraint, and the power constraint, with coefficient
\[
s_m=-(\lambda^e+\eta)\frac{w^1g_m}{\thI_m}-\Delta\rho_m\mu_m .
\]
At optimal multipliers $\lambda^\ast$, strong duality gives $\Pi(R)=\max_x\mathcal L(x,\lambda^\ast;R)$, and for any other loads $R'$, weak duality gives $\Pi(R')\le\max_x\mathcal L(x,\lambda^\ast;R')=\Pi(R)+\sum_ms_m(R'_m-R_m)$. Hence, $s$ is a supergradient of $\Pi$. If $R_m>0$, then $n_m>0$, and \eqref{eq:kkt-n} with equality gives $\Delta\rho_m\mu_m=(g_m/\thI_m)[\nu+(\lambda^e+\eta)(w^0-w^{\mathrm{park}})]$, using $\rho_m/\thS_m=1/\thI_m$. By substitution, we obtain
\[
s_m=-\frac{g_m}{\thI_m}\bigl[\nu+(\lambda^e+\eta)\wfull\bigr]=-\Delta\rho_m\MC_m .
\]

\emph{(iii)} Fix $\nu\ge0$ and set $\eta=0$. For each model, let $a_m(\nu)=\frac12\Lambda_m(1-\MC_m(\nu,0)/\pi_m)^+$ and $n_m(\nu)=R_m/\thI_m+a_m(\nu)/(\Delta\thS_m)$, which makes the capacity constraint bind, and let $\mu_m$ satisfy \eqref{eq:kkt-n} with equality. These choices satisfy \eqref{eq:kkt-n}, \eqref{eq:kkt-a}, and complementary slackness for capacity. The number of accelerators used is $N(\nu)=\sum_mg_mn_m(\nu)$, which is continuous and nonincreasing in $\nu$, strictly decreasing wherever some model sells spot work, and tends to $\sum_mg_mR_m/\thI_m\le G$ as $\nu\to\infty$. If $N(0)\le G$, then $\nu=0$ satisfies the fleet constraint and complementary slackness. Otherwise, there is a unique $\nu>0$ with $N(\nu)=G$ since $N$ is strictly decreasing wherever $N(\nu)>\sum_mg_mR_m/\thI_m$. In both cases, all KKT conditions hold, provided that the power cap is slack at this point, and hence the solution is optimal. The multiplier is unique. For any other $\nu'$, the stationarity conditions determine $a(\nu')$ uniquely, and $N(\nu')\ne G$ would violate either the fleet constraint or complementary slackness.

\emph{(iv)} Each $n_m(\nu)$, and hence $N(\nu)$, is nondecreasing in every $R_{m'}$ and $\Lambda_{m'}$. Since $N$ is nonincreasing in $\nu$, the solution of $N(\nu)=G$, or $\nu=0$, is nondecreasing in every $R_{m'}$ and $\Lambda_{m'}$. Moreover, $\MC_m$ is increasing in $\nu$, and hence, by (i), every spot price is nondecreasing in $\nu$, and every spot volume is nonincreasing in $\nu$ for fixed $\Lambda_m$.

\emph{(v)} With $\nu=\eta=0$, we have $\MC_m=g_m\lambda^e\wfull/(\Delta\thS_m)$, and (i) gives the stated price. \qed

\subsection{Proof of \texorpdfstring{\protect\cref{prop:value}}{Proposition 2} (what a fixed discount loses)}\label{app:pf-value}

(i) If capacity binds at neither price, the profit at price $p$ is $f(p)=(p-c)\Lambda(1-p/\pi)$, where $c$ has no capacity component. This profit is a quadratic with leading coefficient $-\Lambda/\pi$, maximized at $p^\ast=\frac12(\pi+c)$, and hence $f(p^\ast)-f(\bar p)=\frac{\Lambda}{\pi}(\bar p-p^\ast)^2$. Summing over periods with weights $\alpha_t=\Lambda_t/\pi$, the sum $\sum_t\alpha_t(\bar p-p_t^\ast)^2$ is minimized over $\bar p$ at the $\alpha$-weighted mean of $p^\ast_t$, where it equals $(\sum_t\alpha_t)\mathrm{Var}_\alpha(p^\ast)$.

(ii) If capacity $K$ binds at $p^\ast$, the provider sells $K$ at $p^\ast$. At $\bar p<p^\ast$, demand exceeds $K$, and with admissions rationed to $K$, the provider sells the same quantity at the same variable cost and a lower price. Therefore, the difference in profit is $(p^\ast-\bar p)K$. \qed

\subsection{Proof of \texorpdfstring{\protect\cref{lem:fifo}}{Lemma 2} (deadlines as cumulative constraints)}\label{app:pf-fifo}

We order the units of work by deadline, breaking ties by admission time. Since all work has the same relative deadline $D$, this order is the order of admission, and hence first-in-first-out service removes units in this order. Let $X_k=\sum_{j\le k}(x_j+z_j)$ denote the cumulative work removed by the end of step $k$. Under first-in-first-out, the units removed by step $k$ are exactly the first $X_k$ units in the order. The work due by step $k$ consists of the old backlog due by then, $d_k$, plus the work admitted at steps $j$ with $j+D-1\le k$. Every unit due by $k$ is removed by $k$ if and only if $X_k$ is at least this amount, which is the left inequality. The right inequality states that no more work is removed than has arrived. Conversely, any nondecreasing $X$ that satisfies both inequalities for all $k$ is realized by first-in-first-out service of $X_k-X_{k-1}$ units at step $k$. \qed

\subsection{Proof of \texorpdfstring{\protect\cref{thm:defer}}{Theorem 1} (deferral pricing)}\label{app:pf-defer}

\emph{Reformulation.} Let $f_{s\tau}\ge0$ denote the work admitted in period $s$ and served in period $\tau\in\mathcal W_s$. Since work admitted at $s$ and served at $\tau$ is in the backlog at the end of periods $s,\dots,\tau-1$, the holding cost $h\sum_tq_{t+1}$ equals $h\sum_{s,\tau}(\tau-s)f_{s\tau}$ for any assignment. Consider
\begin{equation}\label{eq:defer-flow}
\max_{a,f\ge0}\ \sum_s a_s\pi\Bigl(1-\frac{a_s}{\Lambda_s}\Bigr)-\sum_s\sum_{\tau\in\mathcal W_s}\bigl[c_\tau+h(\tau-s)\bigr]f_{s\tau}\quad\text{s.t.}\ \sum_{\tau\in\mathcal W_s}f_{s\tau}=a_s\ [\lambda_s],\quad\sum_{s:\,\tau\in\mathcal W_s}f_{s\tau}\le K_\tau\ [\mu_\tau].
\end{equation}
Any feasible $(a,f)$ yields aggregate service $x_\tau=\sum_sf_{s\tau}$ that meets every deadline, and therefore the cumulative constraints of \cref{lem:fifo}, at the same profit. Conversely, by \cref{lem:fifo}, any feasible $(a,x)$ of \eqref{eq:defer} is realized by first-in-first-out service, which defines an assignment $f$ with the same profit. Therefore, the two problems have the same optimal value, and we work with \eqref{eq:defer-flow}. Its feasible set is compact since $a_s\le\sum_{\tau\in\mathcal W_s}K_\tau$, and hence an optimum exists.

\emph{KKT conditions.} Stationarity in $a_s$ and $f_{s\tau}$ reads
\[
\pi(1-2a_s/\Lambda_s)-\lambda_s\le0\ \ (=0\text{ if }a_s>0),\qquad \lambda_s-c_\tau-\mu_\tau-h(\tau-s)\le0\ \ (=0\text{ if }f_{s\tau}>0).
\]
The second condition gives $\lambda_s\le C_s$, with equality if $a_s>0$, since then $f_{s\tau}>0$ for some $\tau\in\mathcal W_s$.

(i) If $a_s>0$, then $\pi(1-2a_s/\Lambda_s)=C_s$, and hence $a_s=\frac12\Lambda_s(1-C_s/\pi)$ and $p_s=\frac12(\pi+C_s)$. If $a_s=0$, then $\pi\le\lambda_s\le C_s$.

(ii) By complementary slackness, $f_{s\tau}>0$ only where $\lambda_s=c_\tau+\mu_\tau+h(\tau-s)$, that is, at periods that attain the minimum in \eqref{eq:Ct}.

(iii) If $\mu_\tau=0$ for some $\tau\in\mathcal W_t$, then $C_t\le c_\tau+h(\tau-t)$. When $a_t>0$, (i) gives the bound. When $a_t=0$, the price is the choke price $\pi$, and $\pi\le C_t\le c_\tau+h(\tau-t)$ implies $\pi\le\frac12(\pi+c_\tau+h(\tau-t))$.

(iv) A longer deadline enlarges every window $\mathcal W_s$, and hence the feasible set of \eqref{eq:defer-flow}. \qed

\subsection{Proof of \texorpdfstring{\protect\cref{thm:backfill}}{Theorem 2} (spot backfill)}\label{app:pf-backfill}

Since $\frac{d}{dn}\E\min(R,\thI n)=\thI\bar F(\thI n)$ and $\frac{d}{dn}\E(n-R/\thI)^+=F(\thI n)$, we obtain
\[
\Phi'(n)=\Delta\bigl[(\pi+\beta)\thI\,\bar F(\thI n)+v\thS\,F(\thI n)\bigr]-k=\Delta\bigl[(\pi+\beta)\thI-\bigl((\pi+\beta)\thI-v\thS\bigr)F(\thI n)\bigr]-k .
\]
The assumption $\rho v<\pi+\beta$ is equivalent to $v\thS<(\pi+\beta)\thI$, and hence $\Phi'$ is nonincreasing and $\Phi$ is concave.

(i) Note that $\Phi'(n)+k$ is a convex combination of $\Delta(\pi+\beta)\thI$ and $\Delta v\thS$, and hence $\Phi'(n)\ge\Delta v\thS-k\ge0$.

(ii) If $\Delta\thS v<k$, then $\Phi'(n)\to\Delta v\thS-k<0$ as $n\to\infty$, while $\Phi'(0)=\Delta(\pi+\beta)\thI-k>0$ when $F(0)=0$, by the assumption on $k$. For a load distribution with support on the whole line, such as the normal in (iv), we let $n$ range over $\R$, and $\Phi'(n)\to\Delta(\pi+\beta)\thI-k>0$ as $n\to-\infty$. Since $F$ is continuous, $\Phi'(n^\ast)=0$ for some $n^\ast$, which is optimal by concavity. Solving for $\bar F(\thI n^\ast)$ and dividing the numerator and the denominator by $\thI$ yields \eqref{eq:fractile}.

(iii) We write $\varpi(v)=(A-\rho v)/(B-\rho v)$ with $A=k/(\Delta\thI)$ and $B=\pi+\beta$. The assumption on $k$ is equivalent to $A<B$, and $\varpi'(v)=\rho(A-B)/(B-\rho v)^2<0$. At $v=0$, $\varpi=A/B$.

(iv) If $R\sim N(\hat r,\sigma^2)$, then $\bar F(\thI n^\ast)=\varpi$ gives $\thI n^\ast=\hat r+z^\ast\sigma$. The expected spot capacity is $\Delta\thS\,\E(n^\ast-R/\thI)^+=\Delta\rho\,\E(\thI n^\ast-R)^+=\Delta\rho\sigma\,\E(z^\ast-Z)^+$ with $Z$ standard normal, and $\E(z-Z)^+=z\Phi_N(z)+\phi_N(z)$. Since $z^\ast$ does not depend on $\sigma$, the capacity is proportional to $\sigma$. \qed

\subsection{Proof of \texorpdfstring{\protect\cref{thm:hyst}}{Theorem 3} (placement hysteresis)}\label{app:pf-hyst}

We write \eqref{eq:placement} with loads $u_t\ge0$ and unloads $v_t\ge0$, $n_t-n_{t-1}=u_t-v_t$, and objective $\sum_t[V_t(n_t)-\kappa^+u_t-\kappa^-v_t]$. At an optimum, we may take $u_tv_t=0$ since reducing both by the same amount saves $\kappa^++\kappa^->0$. We attach a free multiplier $\xi_t$ to the $t$-th equality and set $\xi_T=0$. The KKT conditions are
\[
\xi_t\le\kappa^+\ (=\text{ if }u_t>0),\qquad -\xi_t\le\kappa^-\ (=\text{ if }v_t>0),\qquad w_t:=\xi_t-\xi_{t+1}\in\partial V_t(n_t).
\]

(i) Telescoping the last condition with $\xi_T=0$ gives $\xi_t=\sum_{s\ge t}w_s$, and the first two conditions give the stated range and equalities.

(ii) If the replica is loaded at $t_1$ and the next change is an unload at $t_2$, then $\xi_{t_1}=\kappa^+$ and $\xi_{t_2}=-\kappa^-$, and hence $\sum_{s=t_1}^{t_2-1}w_s=\xi_{t_1}-\xi_{t_2}=\kappa^++\kappa^-$. For any window, $\bigl|\sum_{s=t_1}^{t_2-1}w_s\bigr|=|\xi_{t_1}-\xi_{t_2}|\le\kappa^++\kappa^-$.

(iii) By sufficiency of the KKT conditions, the constant plan $n_t\equiv\bar n$, with $u=v=0$, is optimal if there exist $w_t\in\partial V_t(\bar n)$ whose tail sums $\xi_t=\sum_{s\ge t}w_s$ lie in $[-\kappa^-,\kappa^+]$. For $t\notin S$, $0\in\partial V(\bar n)$ since $\bar n$ maximizes $V$. For $t\in S$, the assumption and $V(n)\le V(\bar n)$ give $V_t(n)-V_t(\bar n)\le\delta(n-\bar n)^+$. For $n>\bar n$, this inequality bounds the right derivative of $V_t$ at $\bar n$ by $\delta$. For $n<\bar n$, it shows that the left derivative is nonnegative. Since the superdifferential of $V_t$ at $\bar n$ is the interval between these derivatives, it contains a point $w_t\in[0,\delta]$. With these choices, $\xi_t\in[0,d\delta]\subseteq[-\kappa^-,\kappa^+]$ when $d\delta\le\kappa^+$. If plans must return to $\bar n$, we add the terminal transition from $n_{T-1}$ to $n_T=\bar n$ with its own load and unload costs. Its multiplier $\xi_T$ is then free in $[-\kappa^-,\kappa^+]$ instead of zero. Choosing $\xi_T=-\kappa^-$ gives $\xi_t\in[-\kappa^-,-\kappa^-+d\delta]$, which lies in $[-\kappa^-,\kappa^+]$ when $d\delta\le\kappa^++\kappa^-$. \qed

\subsection{Proof of \texorpdfstring{\protect\cref{rem:pricetaker}}{Remark 1} (price-taking providers)}\label{app:pf-pricetaker}

We replace the revenue $a_s\pi(1-a_s/\Lambda_s)$ in \eqref{eq:defer} by $p^{\mathrm{ex}}_sa_s$ and add the constraints $a_s\le\bar A_s$. Since the reformulation in the proof of \cref{thm:defer} concerns only the service of admitted work, it still applies, and the problem is equivalent to \eqref{eq:defer-flow} with the new revenue and the depth constraints. This problem is a linear program whose feasible set is nonempty (take $a=f=0$) and compact. Therefore, it has an optimum, and the KKT conditions hold with multipliers $\lambda_s$, $\mu_\tau\ge0$, and $\zeta_s\ge0$ for the depth constraints. Stationarity in $a_s$ and $f_{s\tau}$ reads
\[
p^{\mathrm{ex}}_s-\lambda_s-\zeta_s\le0\ \ (=0\text{ if }a_s>0),\qquad \lambda_s-c_\tau-\mu_\tau-h(\tau-s)\le0\ \ (=0\text{ if }f_{s\tau}>0),
\]
and, as in the proof of \cref{thm:defer}, the second condition gives $\lambda_s\le C_s$, with equality if $a_s>0$.

Suppose that $p^{\mathrm{ex}}_s>C_s$. If $a_s<\bar A_s$, then $\zeta_s=0$ by complementary slackness, and the first condition gives $p^{\mathrm{ex}}_s\le\lambda_s\le C_s$, a contradiction. Hence, $a_s=\bar A_s$. Suppose instead that $p^{\mathrm{ex}}_s<C_s$. If $a_s>0$, then $\lambda_s=C_s$, and the first condition with equality gives $p^{\mathrm{ex}}_s=C_s+\zeta_s\ge C_s$, a contradiction. Hence, $a_s=0$. \qed

\subsection{Proof of \texorpdfstring{\protect\cref{lem:pwl}}{Lemma 3} (approximation error)}\label{app:pf-pwl}

(i) The marginal revenues $\pi(1-(2j-1)/K)$ decrease in $j$, and hence an optimal split fills the segments in order. With $w=\hat\Lambda/K$, the value at $a=jw$ is $\pi w\sum_{i=1}^j(1-(2i-1)/K)=\pi w(j-j^2/K)=\pi a(1-a/\hat\Lambda)$, since $\sum_{i\le j}(2i-1)=j^2$. Between breakpoints, $\hat\Pi_K$ is linear, and hence it is the linear interpolation of $\Pi$ at the breakpoints. Since $\Pi$ is concave with $\Pi''=-2\pi/\hat\Lambda$, the interpolation lies below it, and the standard interpolation error bound gives $\Pi(a)-\hat\Pi_K(a)\le\frac12|\Pi''|\,(w/2)^2=\pi\hat\Lambda/(4K^2)$. Divided by $\max\Pi=\pi\hat\Lambda/4$, this bound equals $1/K^2$.

(ii) For $R\sim N(\hat r,\sigma^2)$ and capacity $c$, $\E(R-c)^+=\sigma L(z)$ with $z=(c-\hat r)/\sigma$. Note that $L$ is convex with $L'(z)=-(1-\Phi_N(z))$. Each right-hand side of \eqref{eq:cuts} is $\sigma$ times the tangent of $L$ at $z_i$, evaluated at $z$, and hence it is at most $\sigma L(z)$, with equality at $z=z_i$.

For the gap, fix consecutive grid points $a=z_i<b=z_{i+1}$ and $z\in[a,b]$, and let $s=(L(b)-L(a))/(b-a)$ denote the slope of the chord of $L$ between them. By convexity, $L$ lies below the chord on $[a,b]$, and the chord exceeds the tangents at $a$ and $b$ by $u\,(z-a)$ and $v\,(b-z)$, where $u=s-L'(a)\ge0$ and $v=L'(b)-s\ge0$. Therefore, the gap between $L$ and the larger of the two tangents is at most
\[
\min\bigl\{u(z-a),\,v(b-z)\bigr\}\ \le\ \frac{uv}{u+v}\,(b-a)\ \le\ \frac{u+v}{4}\,(b-a)=\frac{(b-a)\bigl(L'(b)-L'(a)\bigr)}{4},
\]
where the first inequality maximizes the minimum of an increasing and a decreasing linear function of $z$, and the second uses $4uv\le(u+v)^2$ (the gap is zero if $u+v=0$). Since $L'(b)-L'(a)=\Phi_N(b)-\Phi_N(a)\le\phi_N(0)(b-a)$, the gap is at most $\sigma(z_{i+1}-z_i)^2\phi_N(0)/4$. For the grid used, the largest gap over $[-2.5,3]$, computed numerically, is $\cutErr\,\sigma$. \qed

\clearpage
\section{Notation}\label{app:notation}

\begin{table}[htbp]
\centering\small
\begin{tabular}{@{}ll@{}}
\toprule
Symbol & Meaning \\
\midrule
$t$, $\Delta$, $H$ & control period, its length (15 minutes), planning horizon in periods \\
$\Mset$, $m$, $g_m$ & set of models, a model, accelerators per replica \\
$G$, $W_t$ & accelerators in the fleet, power cap (kW) \\
$\tau_m(b)=\tau_{0,m}+\tau_{1,m}b$ & decode-step time at batch $b$ \\
$\sI_m$, $\sS$ & speed floors of the interactive and spot tiers (tokens/s per stream) \\
$\bI_m$, $\bS_m$; $\thI_m$, $\thS_m$ & batch caps; replica throughput at full load, by tier \\
$\rho_m=\thS_m/\thI_m$ & latency exchange rate (\cref{lem:cobatch}) \\
$n_{m,t}$, $u_{m,t}$, $v_{m,t}$, $\tilde n_{m,t}$ & replicas, loads, unloads, effective replicas \\
$L_m$, $\ell_m=L_m/\Delta$, $\kappa^\pm_m$ & load time, its share of a period, load and unload costs \\
$\pi_m$, $p_{m,t}$ & interactive price, spot price (per token) \\
$R_{m,t}$, $\hat r_{m,t}$, $\sigma_{m,t}$ & interactive load (tokens/s), its forecast mean and s.d. \\
$Y_{m,t}$, $O_{m,t}$ & interactive load served and lost \\
$\Lambda_{m,t}$, $a_{m,t}$ & spot demand at zero price, spot work admitted (tokens) \\
$q_{m,t}$, $x_{m,t}$, $z_{m,t}$, $D$ & spot backlog, served, expired; deadline in periods \\
$K_{m,t}$ & spot capacity of a period (tokens), \eqref{eq:spotcap} \\
$\beta_m$, $\gamma_m$, $h_m$ & interactive loss penalty, expiry penalty, holding cost (per token) \\
$w^0$, $w^1$, $w^{\mathrm{park}}$, $\wfull$ & idle, full-load increment and parked power per accelerator; $\wfull=w^0-w^{\mathrm{park}}+w^1$ \\
$e_t$, $\lambda^e_t=e_t\Delta/3600$ & electricity price (\$/kWh), cost of one kW for one period \\
$\nu$, $\eta$, $\mu_m$ & shadow prices of the fleet, power and spot capacity \\
$\MC_m$, $C_t$ & full marginal cost of a spot token \eqref{eq:MC}, effective cost under deferral \eqref{eq:Ct} \\
$K$, $L(z)$ & revenue segments, standard normal loss function \\
\bottomrule
\end{tabular}
\end{table}
\FloatBarrier

\section{Reproducibility}\label{app:repro}

The code is available in \texttt{spot/code/}. The script \texttt{spotsim.py} contains the model, the planning program \eqref{eq:lp}, the forecasts, and the plant. The script \texttt{concept.py} computes the examples of \cref{sec:model,sec:structure,sec:mpc}, writes \cref{tab:params} and the macros quoted in those sections, and draws \cref{fig:batch,fig:static,fig:defer,fig:backfill,fig:hyst}. The script \texttt{run.py} runs every simulation of \cref{sec:numerics}, 510 weeks in all, in about fifteen minutes on a ten-core laptop. The script \texttt{report.py} converts its output into \cref{tab:main} and \cref{fig:days,fig:week,fig:events,fig:sens}, and it writes every number quoted in \cref{sec:numerics}. Linear programs are solved by HiGHS \citep{Huangfu2018} through SciPy. The hour-of-day profile of interactive load is read from \texttt{data/derived\_hod\_profile.json}, which holds the mean share of daily token volume in each hour, computed from 30 days of hourly candles of a public inference marketplace \citep{Qubax2026API}. The tier price ratios quoted in \cref{sec:why} are medians over the tiered endpoints in a snapshot of a model router's public listings taken on 31 October 2026 \citep{OpenRouter2026API}. No other external data are used.